\documentclass[fleqn,usenatbib]{mnras}

\usepackage{newtxtext,newtxmath}

\usepackage[T1]{fontenc}

\DeclareRobustCommand{\VAN}[3]{#2}
\let\VANthebibliography\thebibliography
\def\thebibliography{\DeclareRobustCommand{\VAN}[3]{##3}\VANthebibliography}

\usepackage[dvipdfmx]{graphicx}
\usepackage{amsmath}	
\usepackage{siunitx}
\usepackage{color}
\usepackage{here}
\usepackage{lscape}
\usepackage{comment}

\title[Turbulence profiling with SH sensor at Subaru]{SHARPEST: The atmospheric turbulence profiling experiment using Shack-Hartmann sensor at the Subaru telescope}

\author[H.Ogane et al.]{
    Hajime Ogane$^{1}$\thanks{E-mail: h.ogane@astr.tohoku.ac.jp},
    Yoshito Ono$^{2}$,
    Yosuke Minowa$^{2}$,
    Shin Oya$^{3}$,
    Koki Terao$^{2}$,
    Takumi Akasawa$^{1}$,
    \newauthor
    \ Riki Homan$^{1}$,
    and Masayuki Akiyama$^{1}$
    \\
    $^{1}$Astronomical Institute, Tohoku University, 6-3 Aramaki, Aoba-ku Sendai, Miyagi 980-8578, Japan\\
    $^{2}$Subaru telescope, National Astronomical Observatory of Japan, 650 North Aohoku Place Hilo, HI 96720, USA\\
    $^{3}$National Astronomical Observatory of Japan, 2-21-1 Osawa, Mitaka, Tokyo 181-8588, Japan
}

\date{Accepted XXX. Received YYY; in original form ZZZ}

\pubyear{2023}

\begin{document}
\label{firstpage}
\pagerange{\pageref{firstpage}--\pageref{lastpage}}
\maketitle

\begin{abstract}
Atmospheric turbulence profile plays an important role in designing and operating adaptive optics (AO) systems with multiple laser guide stars. To obtain representative free atmospheric profiles and resolved ground layer profiles for future AO systems at the Subaru telescope, we are conducting the SHARPEST (Shack-Hartmann Atmospheric tuRbulence Profiling Experiment at the Subaru Telescope) project. In this project, we develop a turbulence profiler comprising two Shack-Hartmann (SH) sensors to observe a pair of bright stars through the Subaru telescope with high spatial sampling by 2 cm subapertures. We perform two analyses on the SH spot data: variance analysis on the spot scintillation for free atmospheric profiles, and on the spot slope for ground layer profiles. This paper introduces the initial results of free atmospheric profiles as well as total seeing values and wind profiles obtained by the first two engineering runs. The free atmospheric profiles reconstructed by the two independent SH sensors show good agreement. The results are also consistent with simultaneous measurements by another profiler except for turbulence strength at $\sim1$ km, which is explained by an overestimation problem of scintillation-based profilers. Measured total seeing values are also smaller than the simultaneous measurements, possibly due to the difference in ground layer turbulence between the two sites. The wind profiles show good consistency with the direct measurements by a rawinsonde. Through this study, we establish a method to constrain the free atmospheric profile, the total seeing, and the wind profile by analysing data from a single SH sensor with fine subapertures.
\end{abstract}




\section{Introduction}

The atmospheric turbulence profile, which is represented by the refractive index structure constant as a function of altitude, $C_n^2(h)$, is essential for designing and operating tomographic adaptive optics (AO) systems, which overcomes the cone effect (laser tomography AO; LTAO) or provide larger AO correction field than the isoplanatic angle (ground layer AO; GLAO, multi conjugate AO; MCAO, multi object AO; MOAO) by estimating the three-dimensional structure of wavefront distortion using multiple laser guide stars (LGS) and wavefront sensors (WFS). 
Since the atmospheric turbulence profile depends on observing sites, good knowledge on site is important. 
Besides, it is ideal to measure the profile and update the information for the tomographic wavefront reconstruction in a quasi-real-time of 10-20 minutes (\citealp{gendron2014robustness}, \citealp{farley2020limitations}) because the atmospheric turbulence profile varies on time scales of minutes to hours as well as day-night or seasonal variations.

At the Subaru telescope, two tomographic AO systems are under development: ULTIMATE-START (\citealp{akiyama2020ultimate}, \citealp{terao2022ultimate}), an LTAO system and ULTIMATE-Subaru (\citealp{minowa2022ultimate}), a GLAO system. The LTAO estimates and corrects all the atmospheric turbulence in the cylindrical optical path in a narrow field of view and therefore requires the information of atmospheric turbulence profile from the ground to the highest altitude around 20 km for optimizing the tomographic reconstruction matrix. On the other hand, the GLAO system only corrects wavefront distortion caused by the turbulence at the ground to provide the seeing improvement over a wide field of view. Since the GLAO system is sensitive to the turbulence structure of the ground layer up to $\sim 1$ km, a detailed turbulence profile as fine as a few tens of meters, which resolves turbulence in the telescope dome, is essential to predict the GLAO performance and to design the system.

To date, a number of methods have been proposed and demonstrated to measure atmospheric turbulence profiles in real time using triangulation with two separate stars. Scintillation detection and ranging (SCIDAR; \citealp{rocca1974detection}) is a method for obtaining atmospheric turbulence profiles from covariance maps of scintillation in two directions separated by a few arcseconds. SCIDAR has since been applied in many ways including generalized-SCIDAR (\citealp{avila1997whole}), which enables near-ground profiling by making the detection plane optically conjugated to a negative altitude, and Stereo-SCIDAR (\citealp{shepherd2014stereo}), which improves the signal-to-noise ratio by using separate detectors for each of the two directions. Slope detection and ranging (SLODAR; \citealp{wilson2002slodar}; \citealp{butterley2006determination}) is similar in principle to SCIDAR, but it uses covariance maps of wavefront slopes measured with a Shack-Hartmann (SH) sensor instead of scintillation. Similarly, a method to measure both scintillation values and wavefront slopes using a SH sensor is known as coupled slodar and scidar (CO-SLIDAR; \citealp{vedrenne2007c}). These methods have a tradeoff between the altitude resolution and the highest measurable altitude. Surface layer SLODAR (SL-SLODAR; \citealp{osborn2010profiling}) measures ground layer turbulence profile with a resolution of $\sim$ 10 m by observing a widely-separated star pair but is sensitive only up to $\sim$ 100 m.

The atmospheric turbulence profile is possibly estimated even with a single star from temporal correlation or scintillation spatial scale dependence on the propagation distance. 
Such single-star methods are realized with a small-aperture telescope, while a large-aperture telescope with a diameter larger than $\sim 1$ m is required in the measurements with two stars to measure turbulence at high altitudes. 
Single-star-SCIDAR (\citealp{habib2006single}) is based on the principle of SCIDAR, but measures the correlation of scintillation at different times in a single direction by using the assumption of frozen flow. 
Multi aperture scintillation sensor (MASS; \citealp{tokovinin2007accurate}) is a widely-used method which reconstructs a 6-layer turbulence profile from scintillation measured by multiple annular apertures with different radii.
Since the spatial scale of scintillation depends on the altitude of turbulence, the intensity of the measured scintillation by each annular aperture is converted to an atmospheric turbulence profile. 
The scintillation-based technique of MASS is not sensitive to ground layer turbulence and is often combined with differential image motion monitor (DIMM), which measures total turbulence strength from relative image motion between two separate apertures, to estimate the ground layer turbulence strength with the difference between the MASS and DIMM measurements.

Recently, several new approaches have been proposed to improve the altitude resolution of MASS.
SH-MASS (\citealp{ogane2021atmospheric}) is an applied method that allows MASS to be implemented with a SH sensor.
Thanks to a large number of spatial frequencies extracted from combinations of SH spots, SH-MASS achieves higher altitude resolution than the classical MASS.
Similarly, full aperture scintillation sensor (FASS; \citealp{guesalaga2021fass}) and ring-image next generation scintillation sensor (RINGSS; \citealp{tokovinin2021measurement}) improve altitude resolution by using full aperture image instead of several annular apertures.

Some methods with single star measurements combine scintillation and wavefront slope from a SH sensor. Single CO-SLIDAR (SCO-SLIDAR; \citealp{vedrenne2007c}) is a method that fits the measured scintillation and wavefront slope correlations to a theoretical power spectrum. Shack-Hartmann image motion monitor (SHIMM; \citealp{perera2023shimm}) is a similar procedure, but uses a larger $\sim 4$ cm subaperture. The scintillation-based approach generally has a wide altitude range for the turbulence profiling, while its altitude resolution is determined by the pupil sampling size and is typically limited down to a few hundred meters considering the signal-to-noise ratio for reliable estimation.

In our SHARPEST (Shack-Hartmann Atmospheric tuRbulence Profiling Experiment at Subaru Telescope) project, we characterise atmospheric turbulence profile at the Subaru telescope for the next generation LTAO and GLAO projects. 
We develop a turbulence profiler system that consists of two SH sensors to perform SH-MASS and SLODAR simultaneously.
Ground layer turbulence profile below $\sim 400$ m is obtained by making target separation of SLODAR as large as a few arcminutes for the GLAO system, while free atmospheric turbulence above $\sim 1$ km is estimated by SH-MASS by each SH sensor for the LTAO system.
The profiler is installed to one of the Nasmyth focus of the Subaru telescope so that the SLODAR measurements can include local turbulence components caused by the telescope structure and dome, which possibly account for a large fraction of the ground layer turbulence. 

In this paper, we describe the concept and instrument design of the SHARPEST profiling system and present the first on-sky results obtained at the Subaru telescope. We mainly focus on the results from a single SH sensor.
The results of ground layer turbulence profiles by combining the two SH sensor data will be discussed in the next paper.

This paper is organized as follows. 
In section 2, we present instrumentation of the atmospheric turbulence profiler.
Then, the engineering observation at the Subaru telescope is explained in section 3.
In section 4, we describe the analysis of local slope and scintillation detected by the SH sensors including SH-MASS analysis.
Results of measured total seeing, free atmospheric turbulence profile, and wind profile are shown in section 5.
We discuss our results by comparing them with other measurements in section 6.
Our conclusion is summarized in section 7.


\section{Instrumentation}
\subsection{Design concept}

The purpose of the system is to measure free atmospheric turbulence profiles with a wide altitude range of up to $\sim 20$ km using SH-MASS and ground layer turbulence profiles with a high altitude resolution of a few tens of meters using the SLODAR technique.
To detect scintillation with the SH sensor, subaperture size must be reduced to a few cm scale, as in SCO-SLIDAR (\citealp{vedrenne2007c}) and SHIMM (\citealp{perera2023shimm}).
According to \citet{ogane2021atmospheric}, in which the relation between altitude resolution and the lowest measurable altitude of SH-MASS is studied, a subaperture size of 2 cm is optimal for detecting turbulence at higher than $\sim 1$ km.

The altitude resolution of SLODAR is determined by the subaperture size and angular separation of the observed star pair.
A high altitude resolution of $\sim$ 20 m is achieved by observing a star pair with a separation of $\sim$ 3.5 arcmin.
Although the field of view at the optical-side Nasmyth focus is 3.5 arcmin in diameter, a larger field of view up to $\sim$ 5 arcmin is achieved by allowing partial vignetting of the pupil.
Actually, only a part of the telescope pupil with a size of $1.32 \times 1.32$ m is sufficient considering that the typical spatial scale of scintillation is at most $\sim$ 10 cm and that ground layer below $\sim$ 400 m is measurable by the SLODAR technique.
Each star should be brighter than $\sim 6$ magnitude in the $V$ or $R$ band to achieve a spot signal-to-noise ratio of more than $\sim 5$ with the 2 cm subaperture.
Although the number of targets is very limited due to the angular separation and the brightness, there is at least one target that is observable from the Subaru telescope throughout a year.

The designed specification of the turbulence profiler system is summarized in Table \ref{table:ProfilerSpec}.

\begin{table}
 \caption{Specification of the atmospheric turbulence profiler}
 \label{table:ProfilerSpec}
 \centering
  \begin{tabular}{ll}
   \hline
   Parameter & Designed value\\
   \hline \hline
   \multicolumn{2}{c}{Target}\\
   \hline
   Number of targets & 2\\
   Angular separation & 180-300 arcsec\\
   Magnitude & $<\sim 6$ at $V$ or $R$ band\\
   \hline
   \multicolumn{2}{c}{Telescope}\\
   \hline
   Effective aperture size & 1.32 $\times$ 1.32 m\\
   \hline
   \multicolumn{2}{c}{Turbulence profiler}\\
   \hline
   Number of SH sensors & 2\\
   Number of subapertures & 66 $\times$ 66\\
   Subaperture size & 2.0 cm\\
   FA profiling method & SH-MASS\\
   FA height resolution & a few km\\
   FA height range & $>\sim 1$ km\\
   GL profiling method & SLODAR\\
   GL height resolution & $\sim 20$ m\\
   GL height range & $<\sim 400$ m\\
   \hline
  \end{tabular}
\end{table}


\subsection{Optical and mechanical design}
\label{sec:opticaldesign}

\begin{figure}
    \centering
    \includegraphics[width=\columnwidth]{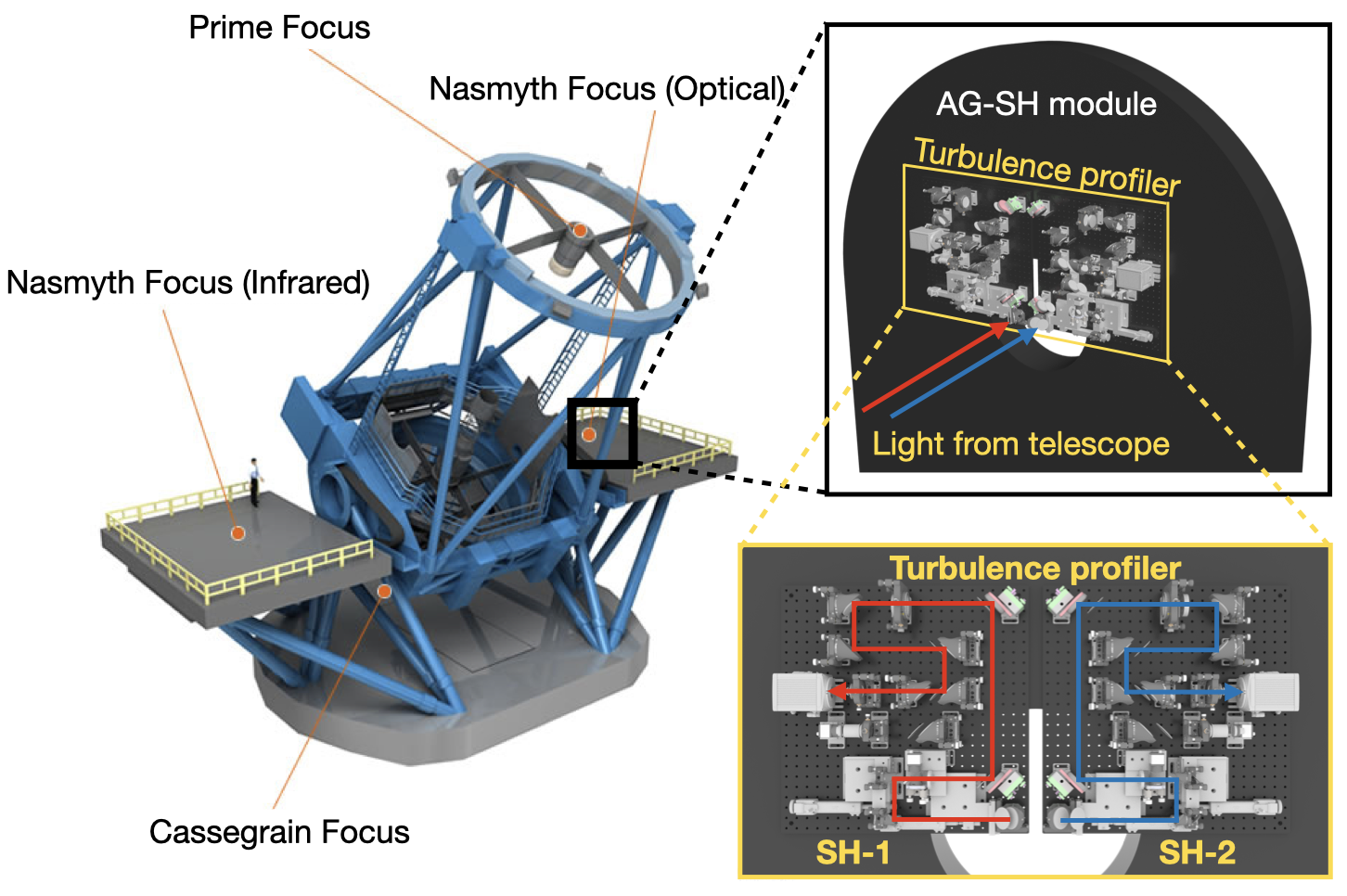}
    \caption[Conceptual drawing of the atmospheric turbulence profiler]{Conceptual drawing of the atmospheric turbulence profiler. The profiler system is installed on the surface of the AG-SH module, which is shown as a black semi-circular object on one of the Nasmyth focus platforms (optical Nasmyth side) of the Subaru telescope. (credit of telescope figure: National Astronomical Observatory of Japan)}
    \label{fig:ProfilerPosition}
\end{figure}

\begin{figure}
    \centering
    \includegraphics[width=\columnwidth]{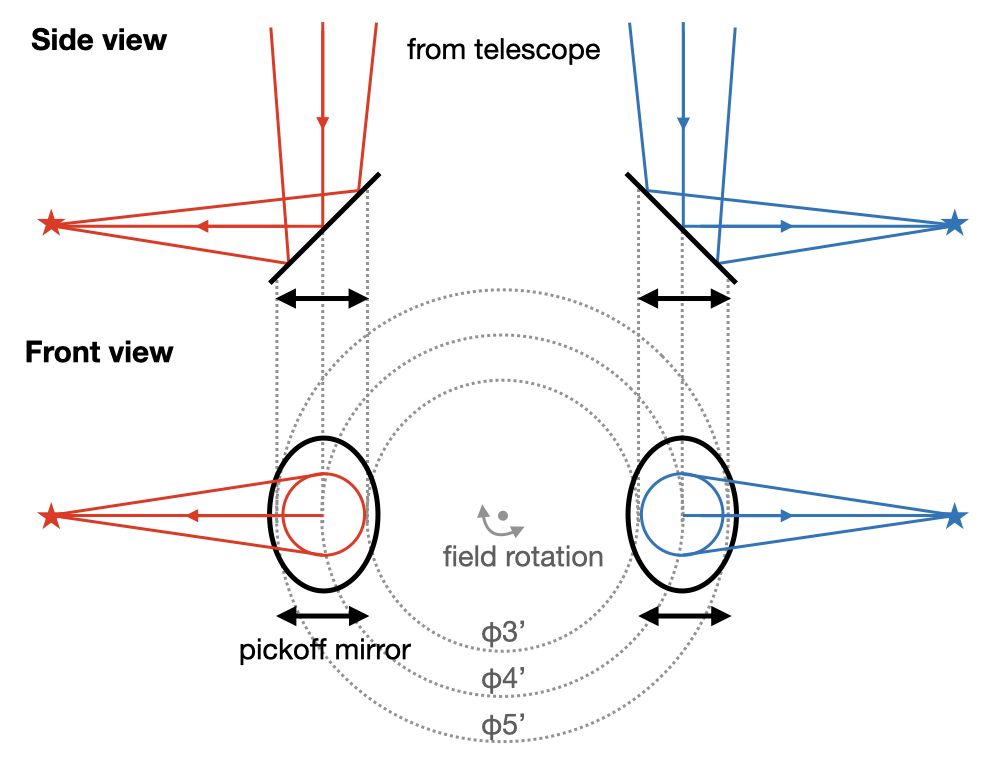}
    \caption[Schematic figure of pickoff position]{Schematic figure of pickoff position. Red and blue lines represent convergent beams of the two target stars through the telescope. Two pickoff mirrors of the turbulence profiler shown as black shape move linearly in a range of 1.5-2.5 arcminutes from the centre of the telescope field of view depending on the star separation. Field rotation is performed by an image rotator of the telescope.}
    \label{fig:ProfilerPickoff}
\end{figure}

Figure \ref{fig:ProfilerPosition} shows a conceptual drawing of the turbulence profiler attached to the Subaru telescope.
The system consists of two symmetric SH sensors and is installed on the surface of the Auto Guider \& Shack-Hartmann module (AG-SH module) at the optical-side Nasmyth focus of the telescope.
We call a SH sensor on the rear side of the telescope as SH sensor 1 (SH-1) and that on the front side as SH sensor 2 (SH-2).
Light from the objects reaches the turbulence profiler as a convergent beam through a telescope composite focal length of $\sim$ 104 m. 
As shown in figure \ref{fig:ProfilerPickoff}, the turbulence profiler uses pickoff mirrors to pick up beams from the two stars at 470 mm upstream of the telescope focal plane and introduce the rays into the respective two SH sensors.
The positions of the two beams are fixed at the pickoff plane during observation thanks to the telescope tracking and the field rotation by the telescope image rotator.
Also, to deal with the different separation angles of observation targets, the pickoff mirrors move in a range of 1.5-2.5 arcminutes from the center of the telescope field of view depending on the star separation. 

\begin{figure}
    \centering
    \includegraphics[width=\columnwidth]{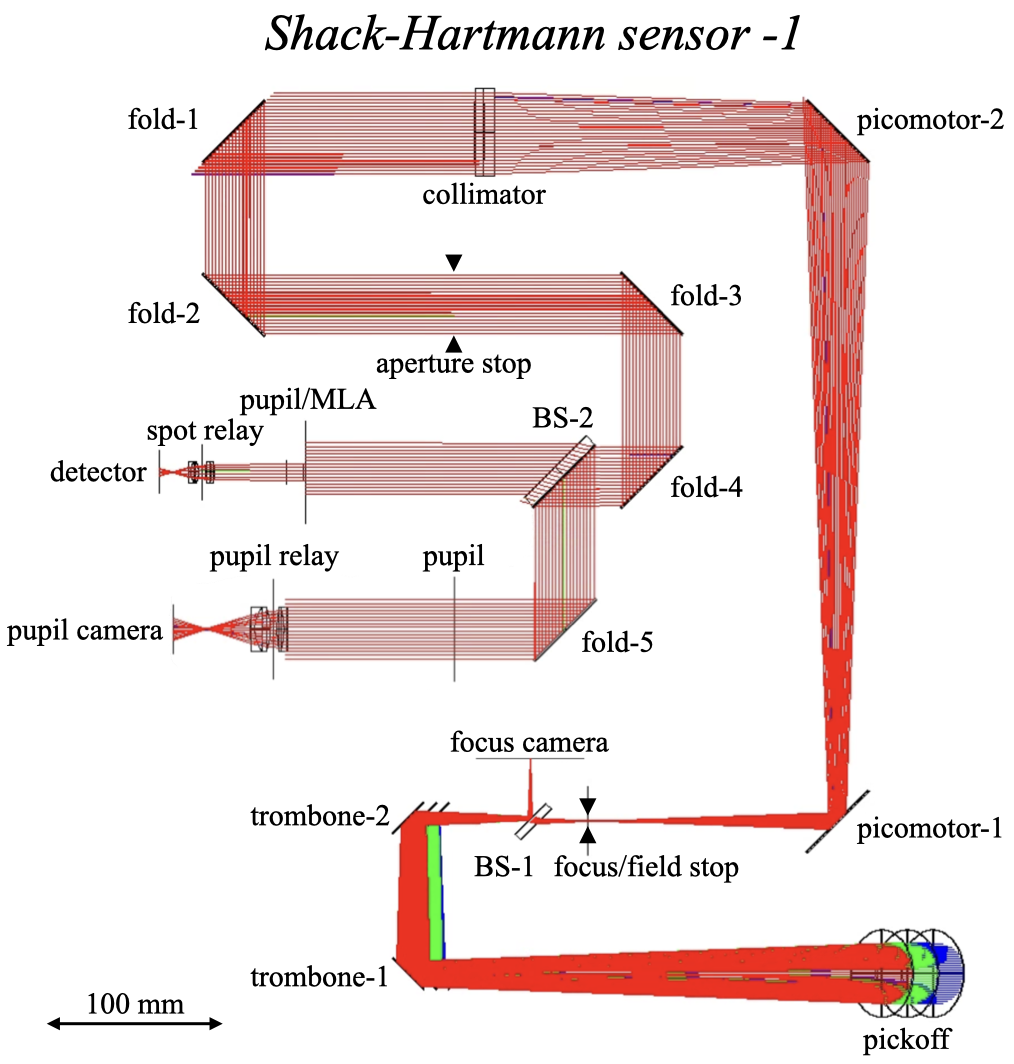}
    \caption[Optical design of the turbulence profiler]{Optical design of the turbulence profiler. The profiler system consists of two symmetrical SH sensors. Only SH sensor 1 is shown in this figure. Rays from the telescope come in the direction from the front to the back of the paper and are introduced into each SH sensor by the pick-off mirrors at the bottom of the figure. Blue, green, and red lines represent rays from two stars with a separation of 3, 4, and 5 arcmin, respectively.}
    \label{fig:ProfilerOptics}
\end{figure}

Figure \ref{fig:ProfilerOptics} shows the details of one of the SH sensors.
After the beam pickoff, a trombone mirror system, which consists of two fold mirrors, adjusts the optical path length to make the focal plane at a fixed position. 
The light is then collimated by a collimator lens with a focal length of 750 mm. 
A microlens array that consists of $66 \times 66$ microlenses with a diameter and pitch of $150\ \mathrm{\mu m}$ and a focal length of 10 mm is arranged at a position conjugated to the telescope primary mirror. 
The pitch of the microlens array corresponds to a subaperture of $2.0\ \mathrm{cm}$ on the primary mirror. 
Thus, the pupil area covered by the microlens array corresponds to a square area of $\sim$ $1.32\times1.32\ \mathrm{m}$ on the primary mirror, which is a portion of the entire aperture of the 8.2 m telescope. 
Spots by the microlens array are then reduced to 40\% in F-number and relayed to the SH sensor detector.
The diffraction limit of 2 cm subaperture determines the full width at half maximum (FWHM) size of each spot of 7.55 arcsec at 600 nm, which is sampled by $\sim 3$ pixels on the detector.

We also have telescope focus and pupil cameras to assist object acquisition to the optical axis of each SH sensor and to check the observed pupil position, respectively.
The focus camera has a field of view of $10 \times 10\  \mathrm{arcsec}$, which allows us to check the tracking by the telescope, field rotation by the image rotator, acquisition by the pick-off arms, and focusing by the trombone system. 
The pupil camera monitors the area of $3\ \mathrm{m}$ diameter on the primary mirror. The observed areas of the telescope pupil are checked by looking at the shade position by telescope structures, such as spiders, or the secondary mirror, on the pupil camera or the SH sensors.
When the observed pupil area is different between the two SH systems, we use two motorized picomotor mirrors to adjust the observed pupil areas to each other. 
The remaining misregistration between the two areas is precisely matched by SLODAR analysis software using spatial synchronization of spots on the two SH sensors.
The details for pupil matching will be discussed in the next paper.
The optical specification of the SH sensor is summarized in Table \ref{table:SHsensorSpec}.

\begin{figure}
    \centering
    \includegraphics[width=\columnwidth]{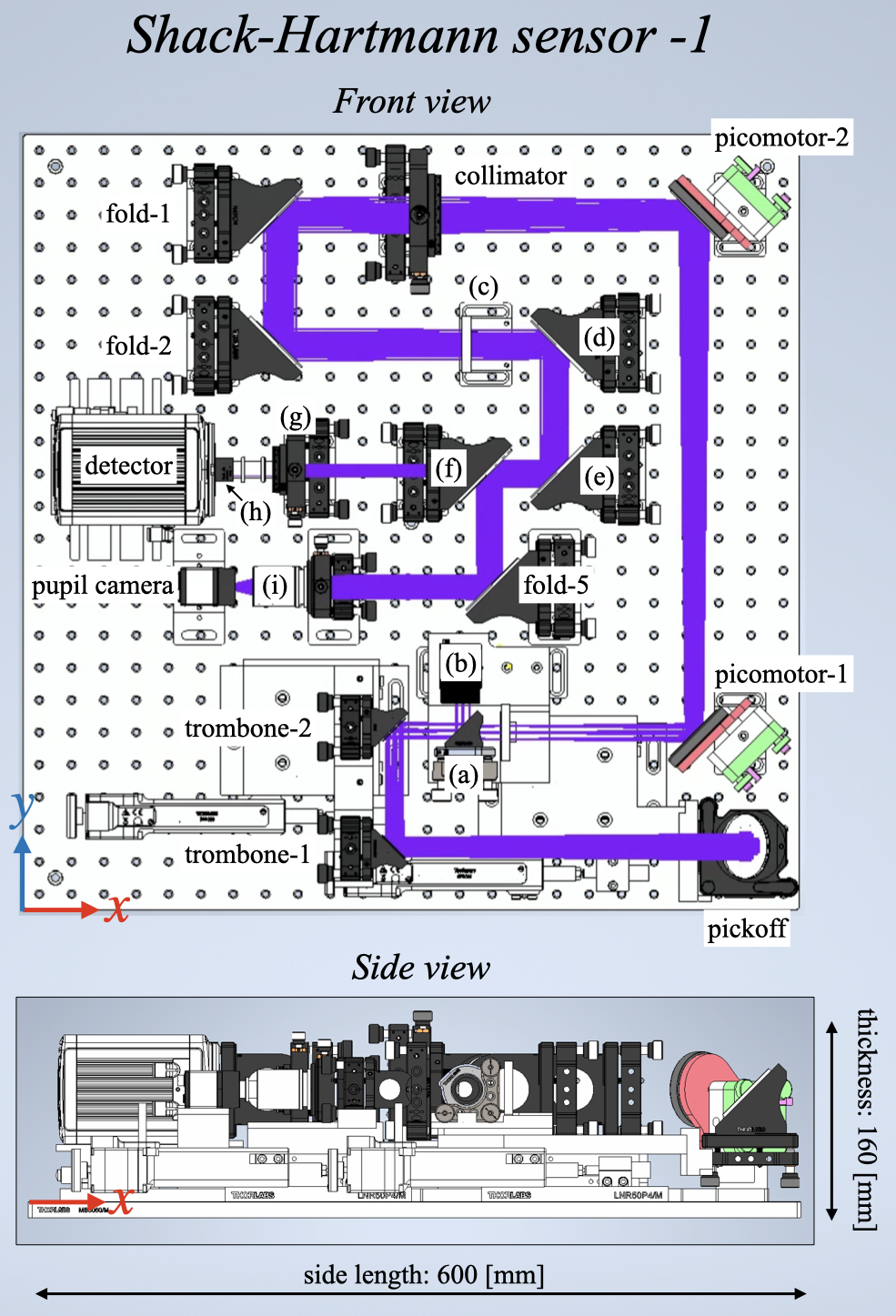}
    \caption[Mechanical layout of the turbulence profiler]{Mechanical layout of the turbulence profiler. Only SH sensor 1 is shown. The other sensor, SH-2, is the symmetrical version of SH-1. The purple lines on the front view show an optical path. Alphabetically labelled optics are as follows: (a)beam sampler-1, (b)focus camera, (c)aperture stop, (d)fold mirror-3, (e)fold mirror-4, (f)beam sampler-2, (g)microlens array, (h)spot relay, (i)pupil relay.}
    \label{fig:ProfilerMechanical}
\end{figure}

Figure \ref{fig:ProfilerMechanical} shows a 3D CAD drawing of the completed mechanical design of the SH sensor 1. 

\begin{table}
 \caption{Specification of the SH sensor}
 \label{table:SHsensorSpec}
 \centering
  \begin{tabular}{ll}
   \hline
   Parameter & Designed value\\
   \hline \hline
   Number of subapertures & 66 $\times$ 66\\
   Subaperture size & 2.0 cm\\
   Subaperture FoV & 25 arcsec\\
   Number of pixels in a subaperture & 10\\
   Pixel scale & 2.5 arcsec $\mathrm{pix^{-1}}$\\
   Pixel size & 6.5 $\mathrm{\mu m}$\\
   Readout noise & 1.6 $\mathrm{electron\ pix^{-1}}$\\
   Quantum efficiency & >50\% at 420-800 nm\\
   & 82\% at peak\\
   \hline
  \end{tabular}
\end{table}

\subsection{Data acquisition system}
\label{sec:das}

We utilize the ORCA-Flash4.0 V2 from Hamamatsu Photonics as detectors for the SH sensors.
This detector is a $2048 \times 2048$ CMOS sensor with a pixel size of $6.5\ \mathrm{\mu m}$, a readout noise of $1.6\ \mathrm{electron\ pix^{-1}}$ in RMS and a quantum efficiency of $\sim 80 \%$ at 600 nm.
The fastest frame rate is $100\ \mathrm{Hz}$ when reading the entire region, but an even faster readout is achieved by limiting the number of readout rows.
In observation, the turbulence profiler reads 800 rows at maximum to cover the entire image of the microlens array.
The fastest frame rate corresponding to the 800-row readout is 230 Hz, but we also use a slower frame rate of 80 Hz for fainter targets.
In addition, data with a faster frame rate of 350 Hz and 700 Hz for brighter targets, by limiting the number of readout rows to 512 and 256 respectively, is also obtained to check the impact of exposure time of data acquisition.

The turbulence profiler synchronizes the image acquisition timing of the two SH sensors by an external trigger signal from a single function generator to the two detectors.
The synchronicity of image acquisition is checked for the obtained data, and it is found that for more than $97\%$ of the frames in the dataset, the timing lag between the two SH sensor acquisitions is within $\pm 50\ \mu s$.
This indicates that synchronization is achieved within $3.5 \%$ of the exposure time even for the data with the shortest exposure time.

The detector works in air-cooling mode. 
We place a partition between the ventilator on the side of the detector and the other optics so that the circulating air does not affect the measurement.
Also, we keep covers at the back side of the detectors open to prevent heat accumulation in the system.

Data acquisition by the detector system, driving of the pickoff and trombone mirror stages, and angle adjustment of the picomotor mirror are controlled by a server computer with Ubuntu 20.04 with a real-time kernel patch and C-based software.


\section{Observations}
\label{sec:observation}

The observation targets of the turbulence profiler are pairs of natural stars with separation angles of 3-5 arcmin. In addition, each star should be brighter than $\sim$ 6 magnitude in the visible wavelength range. There are only five target candidates of the turbulence profiler, but one or more of the five objects are observable at any time of the year. The candidates are summarized in table \ref{table:ProfilerTargets}. 
We also observe a single star as bright as $\sim$ 1 magnitude in $V$ or $R$ band using one of the SH sensors for demonstrating only SH-MASS with high signal-to-noise data.

Engineering observations with the turbulence profiler on board the Subaru telescope have been made twice as of the end of April 2023. The first observation was performed in the first half night of Nov.12, 2022 in Hawaiian standard time (HST). This was the first observation as a turbulence profiler. We confirmed the procedure of target acquisition and performed data acquisition using the target \#1 in Table \ref{table:ProfilerTargets}. Five time-series data are obtained between 23:00 and 24:00 in local time. 
Each time series covered 120 seconds continuously, and the exposure time of the SH sensor was varied among the five data acquisitions depending on the size of the images. For each exposure time setting, we also measured the effect of the sky background and readout noise by taking images at an offset position $\sim 30$ arcsec away from the targets.

The second engineering observation was conducted in the second half night on March 13, 2023 in HST. In addition to data acquisition using target \#4, we observed a double star to determine the pixel scale of the SH sensor and acquired data for SH-MASS using a single bright star. It should be noted that the background measurement at an offset position was basically not performed this time. 
This is because the data analysis of the first engineering observation showed that the readout noise exceeds the Poisson noise of the sky background.
The effect of readout noise was estimated by conducting data acquisition under no illumination inside the dome on March 28, 2023, after the test observation and before the optical system removal.

The data for astronomical and non-astronomical objects acquired during each test observation are summarized in Table \ref{table:AstronomicalObjects} and Table \ref{table:NonAstronomicalObjects}, respectively.

\begin{table*}
 \caption[Targets of turbulence profiler observation]{Target of turbulence profiler observation. The coordinates and magnitudes of the targets are taken from the SIMBAD database (\url{http://simbad.cds.unistra.fr/simbad/}), a: the unit of hour, minute, second, b: the unit of degree, arcmin, arcsec.}
 \label{table:ProfilerTargets}
 \centering
  \begin{tabular}{l|lrrl|l|lrrl}
   \hline
   ID & star A &       &       &       &            & star B &       &       &       \\
   \hline
   & Name   & RA(J2000)    & DEC(J2000)   & mag      & separation & Name   & RA(J2000)    & DEC(J2000)   & mag     \\
   &        & ($\mathrm{hms^a}$)        & ($\mathrm{dms^b}$)        &          & (arcsec)   &        & (hms)        & (dms)        &          \\
   \hline \hline
   1 & HD11727    & 01 55 54.48 & +37 16 40.05 & 5.9(V) & 200.742 & 56 And     & 01 56 09.36 & +37 15 06.60 & 5.0(R) \\
   2 & c Ori     & 05 35 23.16 & -04 50 18.09 & 4.6(V) & 252.228 & 45 Ori     & 05 35 39.48 & -04 51 21.84 & 5.2(V) \\
   3 & ksi Pup   & 07 49 17.66 & -24 51 35.23 & 3.3(V) & 287.568 & 181 Pup    & 07 49 01.67 & -24 54 44.09 & 5.3(V) \\
   4 & alf01 Lib & 14 50 41.17 & -15 59 49.96 & 4.8(R) & 230.532 & alf02 Lib & 14 50 52.71 & -16 02 30.40 & 2.6(R) \\
   5 & eps01 Lyr & 18 44 20.34 & +39 40 12.45 & 4.7(V) & 208.524 & eps02 Lyr & 18 44 22.78 & +39 36 45.79 & 5.2(V) \\
   \hline
  \end{tabular}
\end{table*}

\begin{table*}
 \caption{Data of astronomical objects obtained by the two engineering observations. a: Size of the used region on SH sensor, b: Exposure time of data acquisition, c: Acquisition frequency, d: Duration of data acquisition, e: Synchronization of the acquisition by the two sensors}
 \label{table:AstronomicalObjects}
 \centering
  \begin{tabular}{lllrrrrrrll}
   \hline
   Time & Obj-1 & Obj-2 & Mag-1 & Mag-2 & $\mathrm{Size^a}$               & $\mathrm{Exp.^b}$ & $\mathrm{Freq.^c}$ & $\mathrm{Dur.^d}$  & $\mathrm{Sync.^e}$ & Note \\
   (HST)   &       &       & (mag) & (mag) & ($\mathrm{pix^2}$) & (ms) & (Hz)  & (sec) &       &  \\
   \hline \hline
   Nov.12, 2022 \\
   \hline
   23:17 & HD 11727 & 56 And & 5.9(V) & 5.0(R) & $800 \times 800$ & 12.99 &  77 & 120 & yes & \\
   23:37 & HD 11727 & 56 And & 5.9(V) & 5.0(R) & $800 \times 800$ & 12.99 &  77 & 120 & yes & \\
   23:42 & HD 11727 & 56 And & 5.9(V) & 5.0(R) & $512 \times 512$ &  4.33 & 231 & 120 & yes & \\
   23:45 & HD 11727 & 56 And & 5.9(V) & 5.0(R) & $256 \times 256$ &  2.89 & 347 & 120 & yes & \\
   23:49 & HD 11727 & 56 And & 5.9(V) & 5.0(R) & $512 \times 512$ &  4.33 & 231 & 120 & 1.45 ms lag & \\
   \hline
   Mar.14, 2023 \\
   \hline
   02:42 & Arcturus & & -0.05(V) & & $256 \times 1024$ & 1.43 & 698 & 120 & no & \\
   02:45 & Arcturus & & -0.05(V) & & $256 \times 1024$ & 1.43 & 698 & 120 & no & \\
   03:30 & & Arcturus & & -0.05(V) & $256 \times 1024$ & 1.43 & 698 & 120 & no & \\
   03:34 & & Arcturus & & -0.05(V) & $512 \times 1024$ & 2.87 & 349 & 120 & no & \\
   03:37 & & Arcturus & & -0.05(V) & $800 \times 1024$ & 4.32 & 231 & 114 & no & \\
   04:16 & & zet UMa & & 2.2/3.9(V) & $800 \times 1024$ & 500 & 2 & 10 & no & pixel scale calibration\\
   04:21 & zet UMa & & 2.2/3.9(V) & & $800 \times 1024$ & 500 & 2 & 10 & no & pixel scale calibration\\
   04:44 & alf02 Lib & alf01 Lib & 2.6(R) & 4.8(R) & $512 \times 1024$ &  2.88 & 347 &  82 & yes & \\
   05:07 & alf02 Lib & alf01 Lib & 2.6(R) & 4.8(R) & $800 \times 1024$ & 12.95 &  77 & 124 & 2.85 ms lag & \\
   05:13 & alf02 Lib & alf01 Lib & 2.6(R) & 4.8(R) & $800 \times 1024$ & 12.95 &  77 & 124 & 2.85 ms lag & \\
   05:16 & alf02 Lib & alf01 Lib & 2.6(R) & 4.8(R) & $800 \times 1024$ &  4.32 & 232 &  89 & yes & partly cloudy\\
   05:20 & alf02 Lib & alf01 Lib & 2.6(R) & 4.8(R) & $800 \times 1024$ & 12.95 &  77 &  57 & yes & partly cloudy\\
   \hline
  \end{tabular}
\end{table*}

\begin{table*}
 \caption{Data of non-astronomical objects obtained by the two engineering observations}
 \label{table:NonAstronomicalObjects}
 \centering
  \begin{tabular}{lllrrrrll}
   \hline
   Time (HST) & Obj-1 & Obj-2 & Size               & Exp. & Freq. & \# of frames & Sync. & Note \\
        &       &       & ($\mathrm{pix^2}$) & (ms) & (Hz)  &              &       &      \\
   \hline \hline
   Nov.09, 2022 \\
   \hline
   15:39 &           & dome flat & $800 \times 800$ & 100 &  10 & 5 & no & \\
   15:39 & dome flat &           & $800 \times 800$ & 100 &  10 & 5 & no & \\
   \hline
   Nov.13, 2022 \\
   \hline   
   23:24 & & sky background & $800\times800$ & 12.98 &  77 & 1000 & no & \\
   23:24 & sky background & & $800\times800$ & 12.98 &  77 & 1000 & no & \\
   23:48 & & sky background & $256\times256$ &  2.88 & 347 & 1000 & no & \\
   23:48 & sky background & & $256\times256$ &  2.88 & 347 & 1000 & no & \\
   23:51 & & sky background & $512\times512$ &  4.32 & 231 & 1000 & no & \\
   23:52 & sky background & & $512\times512$ &  4.32 & 231 & 1000 & no & \\
   \hline
   Mar.13-14, 2023 \\
   \hline
   20:41 & dome flat      &                & $800\times1024$ &  100 &  10 &   10 & no & \\
   20:42 &                & dome flat      & $800\times1024$ &  100 &  10 &   10 & no & \\
   \hline
   Mar.28, 2023 \\
   \hline
   18:39 & bias & & $256\times1024$ &  1.43 & 698 & 1000 & no & \\   
   18:39 & & bias & $256\times1024$ &  1.43 & 698 & 1000 & no & \\   
   18:41 & bias & & $512\times1024$ &  2.88 & 349 & 1000 & no & \\   
   18:41 & & bias & $512\times1024$ &  2.88 & 349 & 1000 & no & \\   
   18:42 & bias & & $800\times1024$ &  4.32 & 232 & 1000 & no & \\   
   18:42 & & bias & $800\times1024$ &  4.32 & 232 & 1000 & no & \\   
   18:44 & bias & & $800\times1024$ & 12.95 &  77 & 1000 & no & \\   
   18:44 & & bias & $800\times1024$ & 12.95 &  77 & 1000 & no & \\   
   \hline
  \end{tabular}
\end{table*}


\section{Analysis methods}
\subsection{Shack-Hartmann spot detection}
\label{sec:spotdetection}

We first define a background frame for each dataset by averaging all the frames obtained without the object.
Then, we define a spot reference position by applying centre-of-gravity (CoG) measurement for an averaged frame of all the background-subtracted object frames.
Since spots are separated by 10 pixels from their neighbours by design, we define a subaperture area on the detector as $9\times9\ \mathrm{pix}$ region centred at the spot reference position.

Once the subaperture areas are determined for each dataset, spot CoG and brightness measurement are applied for each subaperture in each background-subtracted frame.
We simply define spot brightness as the sum of the pixel counts in each subaperture, while we performed the spot CoG measurement with windowing and thresholding to mitigate the noise effect.
The window is set to an area of $5\times5$ pix centred at the spot reference position, considering that the diffraction-limited spot size of the 2 cm subaperture is $3\ \mathrm{pix}$ in FWHM and that the typical seeing at Maunakea is sub-pixel on the detector.
As thresholding, we set a moderate threshold of 1.3-1.5$\sigma$ of detector readout noise level depending on the dataset because a higher threshold results in failure of spot detection when the object spot is fainter due to scintillation.
Instead of setting the moderate thresholding, we impose a slope-detection condition, in which we only trust CoG measurements for subapertures that have more than three pixels satisfying the threshold in the window region.
In the slope analysis using the measured CoG, which is described in section \ref{sec:ana-slopecov}, we only use the CoG of spots that satisfy the condition.
 
\begin{figure}
    \centering
    \includegraphics[width=\columnwidth]{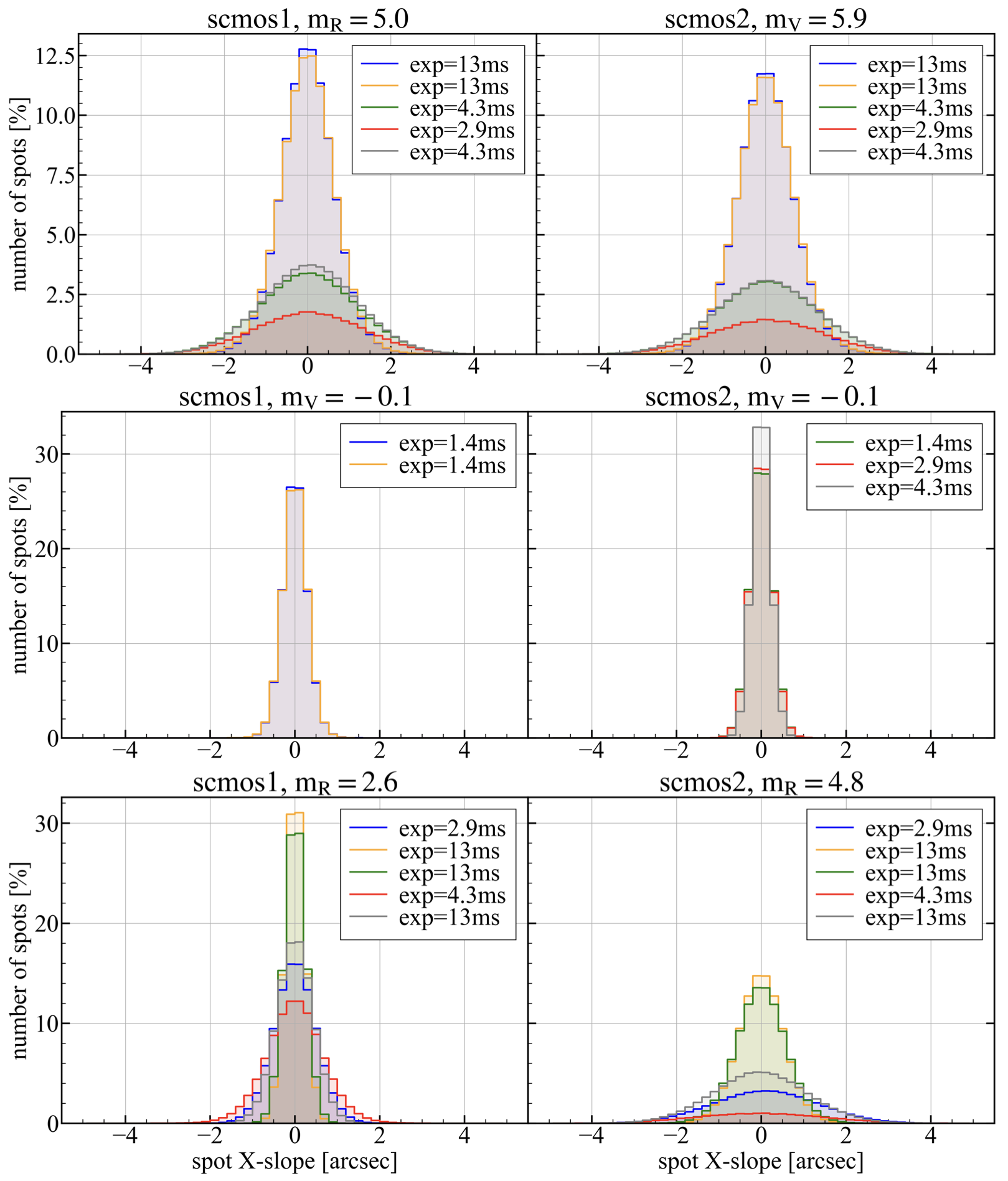}
    \caption[Histogram of detected slope position of spots]{Histogram of detected slope or CoG position of spots in the unit of arcsec. Slope detected by SH sensor 1 (left) and by SH-2 (right) in 23:00-24:00 on Nov.12, 2022 (top), in 2:40-3:40 on Mar.14, 2023 (middle), and in 4:30-5:30 on Mar.14, 2023 (bottom). Only spots that satisfy the slope-detection condition are included. Different colours represent different datasets, i.e. different object brightness and exposure time. Only slope in the x-direction is shown as representative, but similar histograms are obtained for the y-direction.}
    \label{fig:SlopeStatistics}
\end{figure}

A histogram of the measured CoG of all the detected spots in each dataset is shown in figure \ref{fig:SlopeStatistics}. 
The histogram contains only spots that satisfy the slope-detection condition. 
The distribution of the CoG position shows a trend that the dispersion increases as the exposure time decreases. 
The large dispersion is thought to indicate a larger CoG error due to the readout noise.
In slope analysis, the correlations of CoG position are computed among different spots. Hence the readout noise shows a correlation of zero and has little impact on correlation-based analysis. 

\begin{figure}
    \centering
    \includegraphics[width=\columnwidth]{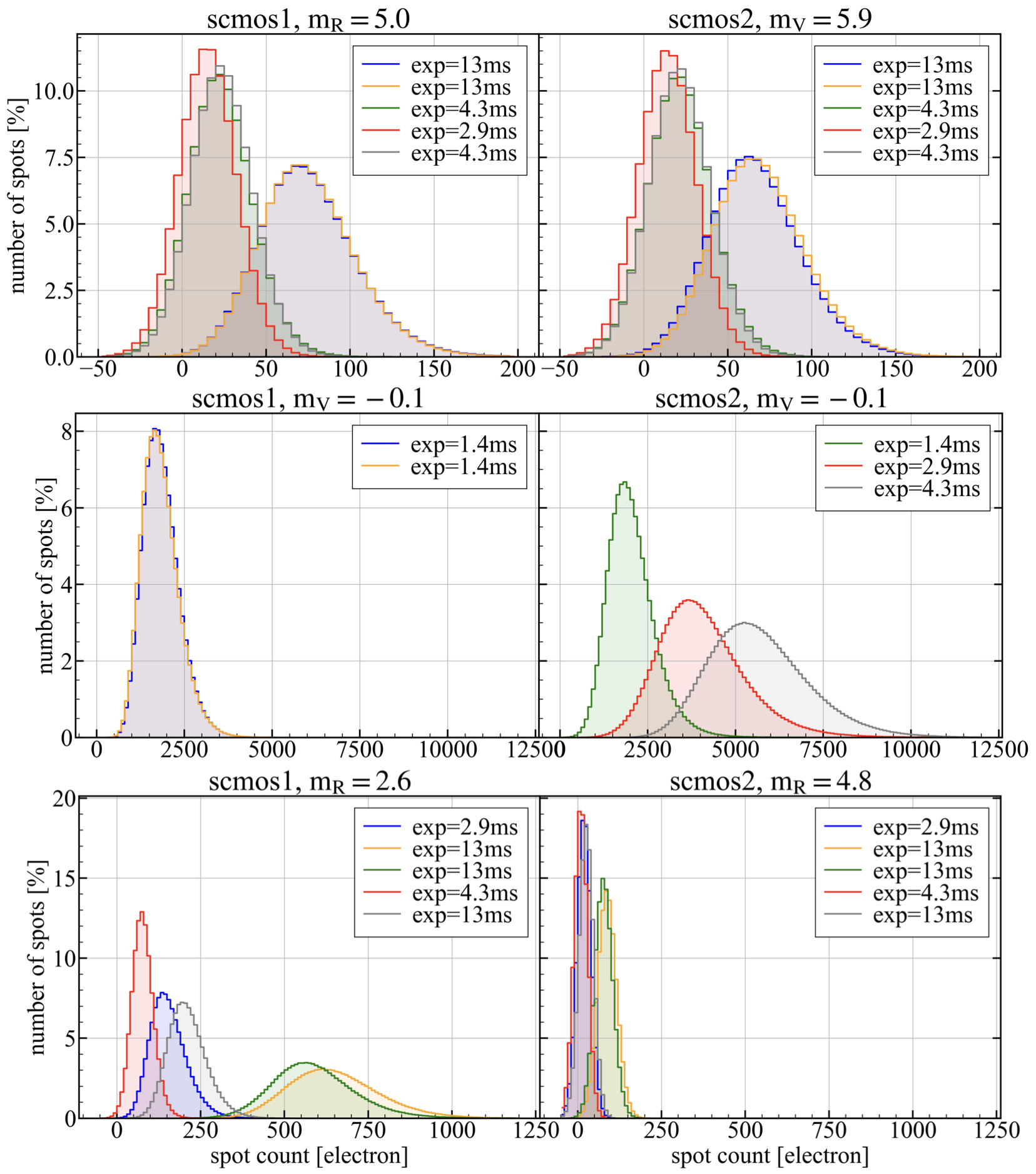}
    \caption[Histogram of detected brightness of spots]{Histogram of detected brightness of spots in the unit of electron. Spot brightness detected by SH sensor 1 (left) and by SH-2 (right) at 23:00-24:00 on Nov.12, 2022 (top), at 2:40-3:40 on Mar.14, 2023 (middle), and at 4:30-5:30 on Mar.14, 2023 (bottom). All the spots are included whether or not they satisfy the detection condition. Different colours represent different datasets, i.e. different object brightness and exposure time.}
    \label{fig:CountStatistics}
\end{figure}

A histogram of the spot brightness of all the spots in each dataset is shown in figure \ref{fig:CountStatistics}. 
Due to the readout noise effect, some spot brightness values are negative when the object is relatively faint or the exposure time of the SH sensor is short. 
In scintillation analysis, the effect of readout noise must be corrected since it is necessary to measure the autocorrelation of spot brightness. 
The correction method is discussed in detail in section \ref{sec:ana-shmass}.

\subsection{Slope auto-covariance}
\label{sec:ana-slopecov}

Total seeing of the atmospheric turbulence is estimated by fitting the observed slope auto-covariance map with an analytically calculated one assuming a turbulence model (\citealp{butterley2006determination}).
``Auto-covariance'' here means covariance between slopes in a single SH sensor, while ``Cross-covariance'' means covariance between slopes in different sensors and is analysed to obtain turbulence profile as SLODAR method (\citealp{wilson2002slodar}, \citealp{butterley2006determination}).
In this analysis, however, we focus on data obtained with a single SH sensor and the SLODAR analysis will be presented in a future paper.

Let the centre position of the $i$-th subaperture on the pupil and corresponding slope measurement in the x-direction be $\boldsymbol{r}_i$ and $s^x_i$, respectively.
The x-slope auto-covariance map $A_x$, which shows the x-slope covariance of two spots as a function of their distance vector $\boldsymbol{r}$, is calculated as 
\begin{align}
A_x(\boldsymbol{r}) = \frac{\sum_{\mathrm{valid}\ i,j}<(s^x_{i}-\overline{s^x})(s^x_{j}-\overline{s^x})>}{\sum_{\mathrm{valid}\ i,j} 1},
\end{align}
where $<\cdot>$ represents the statistical mean of the physical quantity and the sum is performed for valid spot pairs which satisfies $\boldsymbol{r}_j - \boldsymbol{r}_i = \boldsymbol{r}$.
When the covariance is computed, the average of all the slopes, $\overline{s^x}$, is subtracted to remove the overall tip-tilt component which might be induced by non-atmospheric effects such as telescope tracking error or vibration of the optical systems.

Another concern of the non-atmospheric effect is the effect of exposure time.
Since the turbulence profiler has a small subaperture size of 2 cm, turbulence layers move more than the subaperture size during the exposure time depending on the wind speed. 
For example, under the wind speed of 10 $\mathrm{m\cdot s^{-1}}$, the layer moves by 2.9 cm, 4.4 cm, and 12.5 cm in one frame with 350 Hz, 230 Hz, and 80 Hz, respectively. 
Due to the effect, slope measurement along the wind direction might be underestimated and the shape of the auto-covariance signal is smeared or extended in the wind direction.
To minimize the effect, we rotate the coordinate of slope detection to make the wind direction align with the y-axis and use a cross-section of the x-slope auto-covariance map on the x-axis for further analysis.

\begin{figure}
    \centering
    \includegraphics[width=\columnwidth]{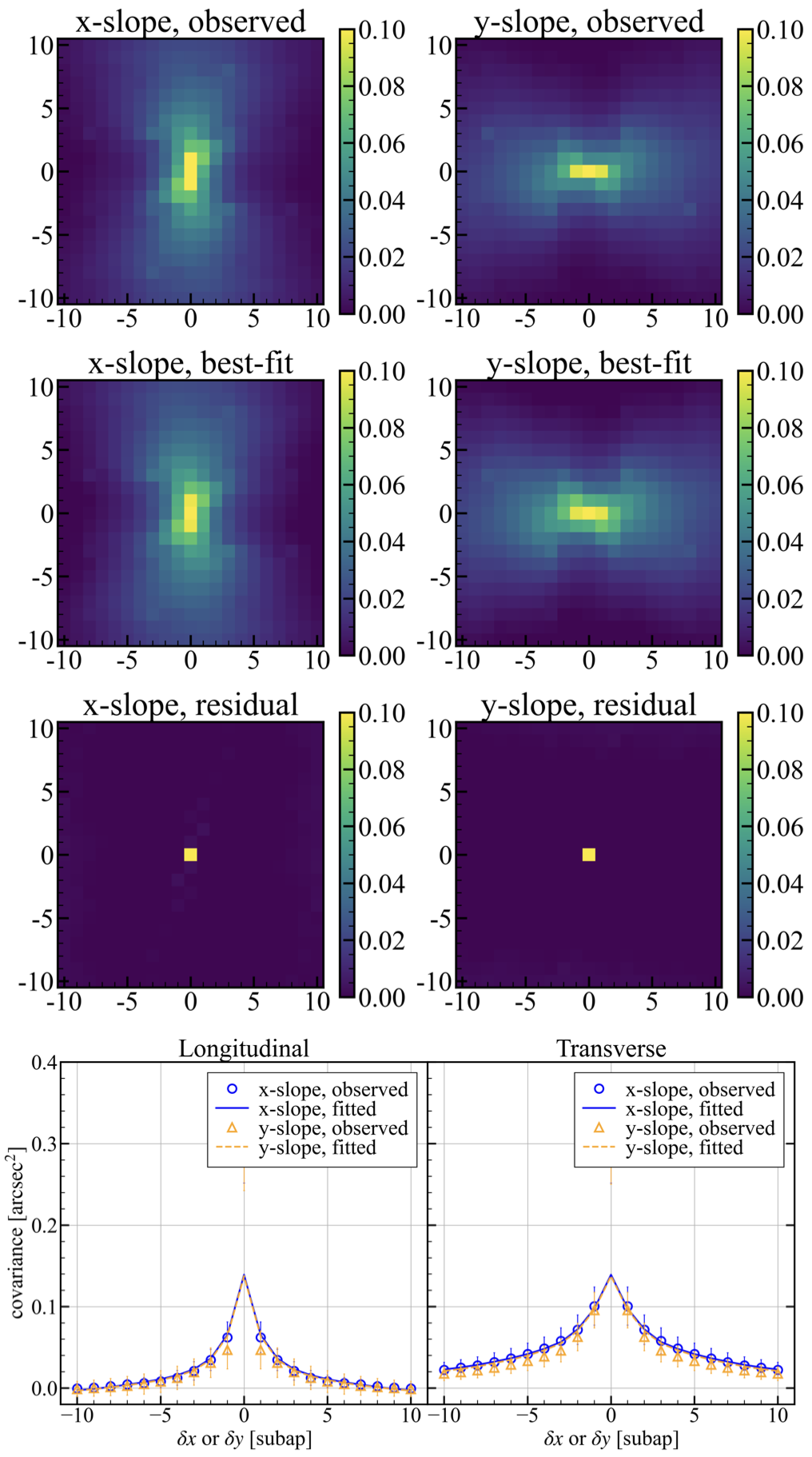}
    \caption[Example of slope auto-covariance analysis]{Example of slope auto-covariance analysis. Coloured maps: Observed auto-covariance maps, best-fit models, and their residuals of slope in x and y directions are arranged from top to bottom. The maps are rotated so that the dominant wind direction should be in the y-direction of the maps.
    Line charts: cross-section in a plane through the centre of the auto-covariance map. The longitudinal profile is a cross-section in $\delta y = 0$ for the x-slope and $\delta x = 0$ for the y-slope, while the transverse profile is a cross-section in $\delta x = 0$ for the x-slope and $\delta y = 0$ for the y-slope.}
    \label{fig:SlopeAutocov}
\end{figure}

The top panels in figure \ref{fig:SlopeAutocov} show an example of the auto-covariance map calculated from the observed spot slope.
We only show auto-covariance within $\pm 10$ subapertures from the centre to remove noisy covariances due to the small number of corresponding subaperture pairs in some datasets. 
The auto-covariance maps are rotated to minimize the exposure effect and the most dominant wind direction corresponds to the y-direction of the maps.

To calculate the analytical model of slope auto-covariance, we adopt an approximated method for accelerated calculation (\citealp{martin2012temporal}, \citealp{ono2016statistics}).
Assuming that the measured slope is the phase difference between two points at the edge of the subaperture, the slope $s^x_i$ is modelled as follows.
\begin{align}
    s^x_{i} = \frac{1}{d}\left[\phi(\boldsymbol{r}_{i}+\frac{d}{2}\boldsymbol{e}_x) - \phi(\boldsymbol{r}_{i}-\frac{d}{2}\boldsymbol{e}_x)\right].
\end{align}
where $d$ is the length of one side of the subaperture, $\phi$ is the phase, and $\boldsymbol{e}_x$ is the unit vector in the $x$ direction. Then the slope covariance of $i$-th and $j$-th spots is written as follows.
\begin{align}
    \label{eq:auto-covariance}
    <s^x_{i}s^x_{j}> = &\frac{1}{2d^2}[-2D_\phi(\boldsymbol{r}_{j} - \boldsymbol{r}_{i}) +D_\phi(\boldsymbol{r}_{j} - \boldsymbol{r}_{i}+d\boldsymbol{e}_x)\nonumber \\ &+D_\phi(\boldsymbol{r}_{j} - \boldsymbol{r}_{i}-d\boldsymbol{e}_x)].
\end{align}
where $D_\phi(\boldsymbol{r})=<[\phi(\boldsymbol{x})-\phi(\boldsymbol{x}+\boldsymbol{r})]^2>$ is the structure function of phase and in the Kolmogorov turbulence model,
\begin{align}
    D_\phi(\boldsymbol{r}) = 6.88 \left(\frac{|\boldsymbol{r}|}{r_0}\right) ^ {5/3},
\end{align}
where $r_0$ is the Fried parameter.
We compute the analytical auto-covariance map for each dataset from equation (\ref{eq:auto-covariance}) and the actual distances of the spots detected in the dataset.

The analytical auto-covariance model is fitted to the measured one to obtain the Fried parameter.
For the fitting, we only use a cross-section of the x-slope auto-covariance map in the x-axis considering the effect of the exposure time.
We also exclude the central pixel of the observed auto-covariance map from fitting since the central value, i.e. variance of the slope contains readout noise variance.

One example of the fitted auto-covariance maps and residuals from the observed maps is shown in the second and third row of figure \ref{fig:SlopeAutocov}, respectively.
The bottom panels show the cross-section of the maps.
We confirmed that the central value of the residual map, which corresponds to readout noise variance, decreases as the signal-to-noise ratio of the measurement or time-averaged spot count is higher.

\subsection{SH-MASS}
\label{sec:ana-shmass}

Free atmospheric turbulence profile is estimated by applying the SH-MASS (\citealp{ogane2021atmospheric}) analysis to the scintillation spatial pattern measured from the intensity fluctuation of the spots on the SH sensor.

In SH-MASS, an aperture to compute scintillation is defined as a pair of two spots on the SH sensor.
Therefore, intensity measured by an aperture with an index of $X$, denoted by $I_X$, is equal to the sum of the intensities measured by two spots on the SH sensor consisting of the aperture $X$.
To characterize scintillation, two indices: normal scintillation index (NSI) and differential scintillation index (DSI) are generally used.
The NSI measured by the aperture $X$, $s_\mathrm{X}$, is defined as
\begin{align}
    s_\mathrm{X}     &= \mathrm{Var}\left[ \frac{I_\mathrm{X}}{\langle I_\mathrm{X} \rangle} \right],
    \label{eq:nsi}
\end{align}
The $<>$ denotes the statistical mean and $\mathrm{Var}$ denotes the statistical variance. 
Similarly, DSI measured by two apertures $X$ and $Y$ is defined as
\begin{align}
    s_\mathrm{XY} &= \mathrm{Var}\left[ \frac{I_\mathrm{X}}{\langle I_\mathrm{X}\rangle}-\frac{I_\mathrm{Y}}{\langle I_\mathrm{Y}\rangle } \right]
    = s_\mathrm{X} + s_\mathrm{Y} - 2\mathrm{Cov}\left[ \frac{I_\mathrm{X}}{\langle I_\mathrm{X}\rangle },\frac{I_\mathrm{Y}}{\langle I_\mathrm{Y}\rangle } \right].
    \label{eq:dsi}
\end{align}

In our analysis, a single atmospheric turbulence profile is obtained from $5\times5$ spots detected in each dataset.
In other words, the apertures for measuring the scintillation indices are limited to combinations of spots separated by a diagonal distance of the $5\times5$ subapertures.
This implicitly assumes that the scintillation correlation at spatial scales larger than $\sim 14\ \mathrm{cm}$, which corresponds to scintillation made by turbulence at a height of $\sim 30$ km, is not considered.
We extract four different spot regions which include $5\times5$ spots from the acquired dataset and perform four independent SH-MASS at the same time on a single SH sensor. 
The number of calculated scintillation indices that are defined by the subapertures of $5\times5$ is 43, including 15 corresponding to the normal scintillation indices and 28 corresponding to the differential scintillation indices.

\begin{figure}
    \centering
    \includegraphics[width=\columnwidth]{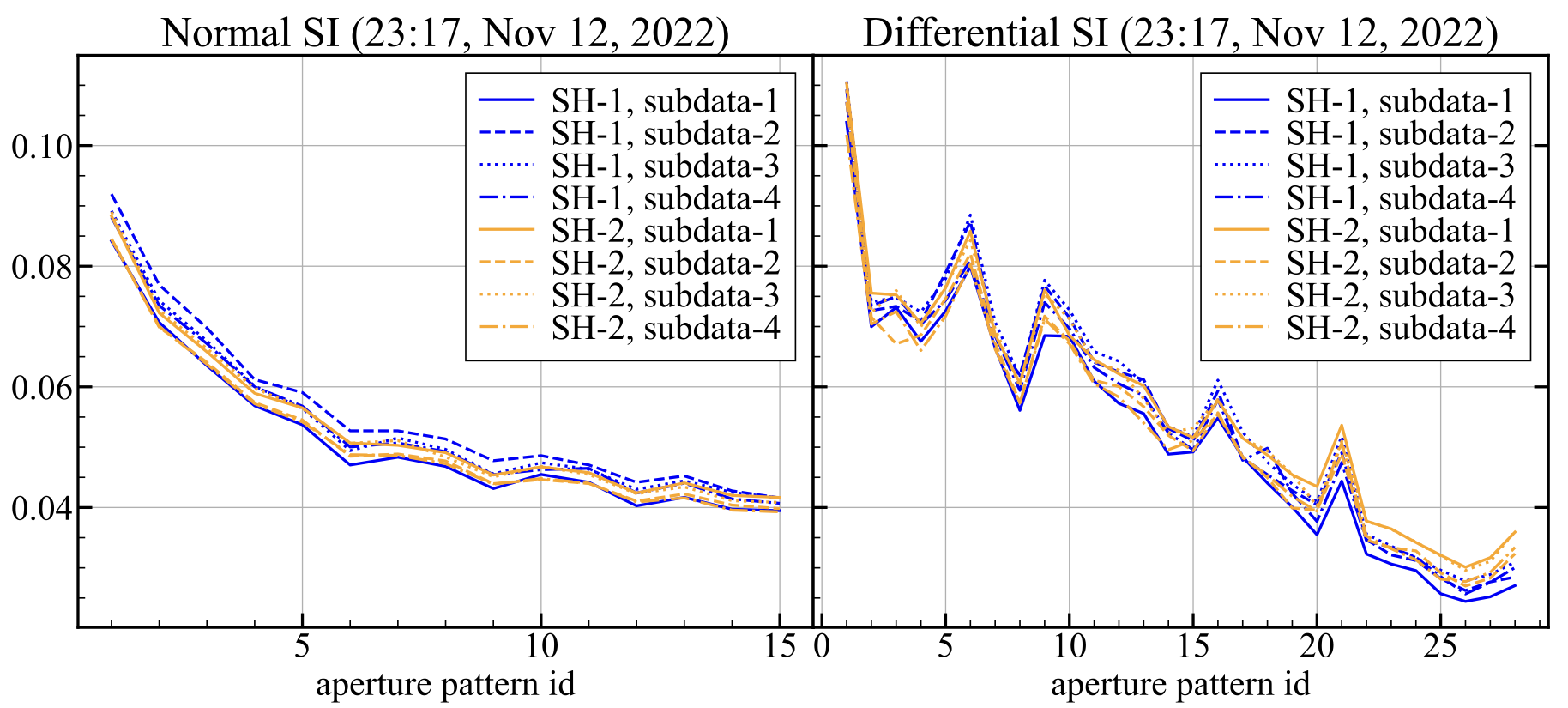}
    \caption[Example of measured scintillation indices]{Example of calculated scintillation indices for the data taken at 23:17, Nov.12, 2022. Normal and differential scintillation index is shown as a function of aperture pattern ID in the left and right panels, respectively. The IDs are given in the order of the distance between the two spots that constitute the aperture. Differences in colours show differences in sensors. Differences in line styles correspond to differences in the subdata of each dataset.}
    \label{fig:NormalScintillationIndex}
\end{figure}

The 43 scintillation indices are calculated from the mean, variance, and covariance of the brightness of each spot.
Since the variance of spot brightness contains photon noise and readout noise components, we correct these components.  
The photon noise is corrected by subtracting the mean brightness of each spot from the variance of the spot based on the fact that photon noise follows a Poisson distribution.
To estimate the readout noise, we perform the same procedure as spot detection on the background datasets.
Because the object is offset in the background datasets, the measured brightness of each subaperture is dominated by the readout noise.
By subtracting the readout noise from the variance of spot brightness, we correct the readout noise effect.
The noise-corrected NSI computed from four different subsets are shown in figure \ref{fig:NormalScintillationIndex}. 
Since these eight results are simultaneous and independent measurements, variations in these lines can correspond to errors in measurement. 
In the following, the standard deviation of the measurement among different subdata is used as the error associated with the scintillation index measurement.

These normal and differential scintillation indices are analytically expressed, assuming an atmospheric turbulence model. Analytical scintillation indices are written as a linear combination of the contribution from multiple turbulence layers at different altitudes:
\begin{align}
    s_\mathrm{X} &= \sum_{i}^{N_\mathrm{layer}} W_{\mathrm{X},i} C_n^2(h_i)\Delta h_i,
    \label{eq:nsi=wj}\\
    s_\mathrm{XY} &= \sum_{i}^{N_\mathrm{layer}} W_{\mathrm{XY},i} C_n^2(h_i)\Delta h_i,
    \label{eq:dsi=wj}
\end{align}
where $h_i$, $C_n^2(h_i)$, and $\Delta h_i$ denote height, refractive index structure constant, and thickness of $i$-th turbulence layer, respectively. 
$N_\mathrm{layer}$ is a total number of turbulence layers.
$W_{\mathrm{X},i}$ and $W_{\mathrm{XY},i}$ are coefficient matrices which represent relationships between each scintillation index and turbulence strength at each altitude, called weighting functions.
Assuming Kolmogorov's turbulence model, weak turbulence approximation, and Fresnel's propagation, the weighting function is written as follows:
\begin{align}
    \label{eq:nwf}
    W_{\mathrm{X},i} &= \iint 1.53f^{-\frac{11}{3}} \left\{ \frac{\sin(\pi\lambda z_i f^2)}{\lambda} \right\}^2 |\mathcal{F}[A_\mathrm{X}(\boldsymbol{x})]|^2 d\boldsymbol{f},\\
    \label{eq:dwf}
    W_{\mathrm{XY},i} &= \iint 1.53f^{-\frac{11}{3}} \left\{ \frac{\sin(\pi\lambda z_i f^2)}{\lambda} \right\}^2 |\mathcal{F}[A_\mathrm{X}(\boldsymbol{x})-A_\mathrm{Y}(\boldsymbol{x})]|^2 d\boldsymbol{f}.
\end{align}
$f=|\boldsymbol{f}|$, $\lambda$ is the observed wavelength, and $z_i$ is distance to the $i$-th turbulence layer.
$A(\boldsymbol{x})$ is the normalized aperture function, which returns 1 divided by the aperture area for $\boldsymbol{x}$ inside the aperture and 0 for others. 
$\mathcal{F}$ denotes Fourier transformation. 

\begin{figure}
    \centering
    \includegraphics[width=\columnwidth]{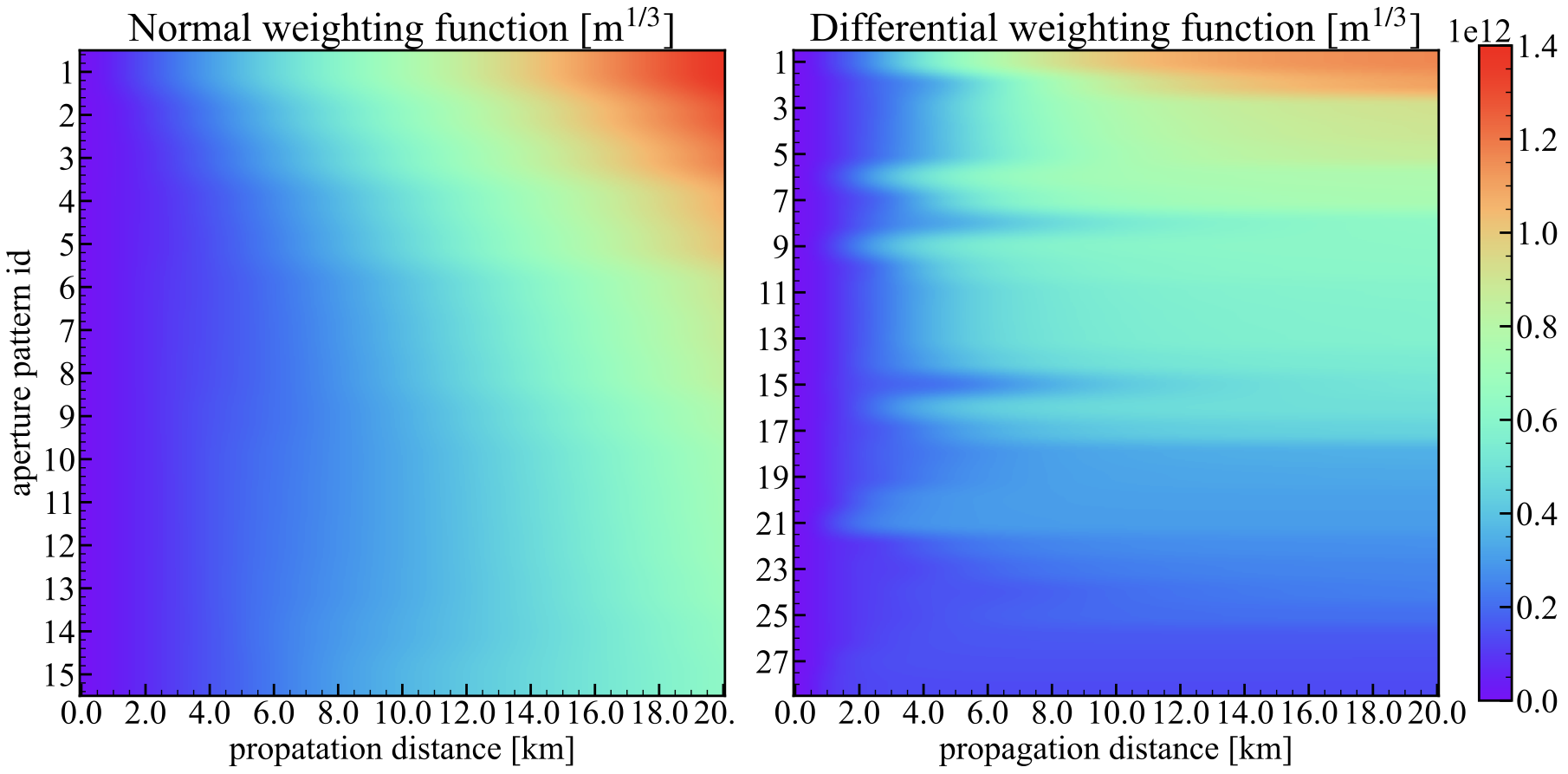}
    \caption[Weighting function matrix]{Weighting function matrix for the SH-MASS analysis. The left panel is for normal aperture patterns and the right panel is for differential aperture patterns. The colour represents the sensitivity of each aperture pattern to scintillation created by the propagation of each distance.}
    \label{fig:WeightingFunction}
\end{figure}

The weighting function matrices that correspond to the 43 scintillation indices are calculated based on equations (\ref{eq:nwf}) and (\ref{eq:dwf}). The aperture function $A_X(\boldsymbol{x})$ is expressed for all the apertures with a size of $1024 \times 1024$ and a sampling of $2\ \mathrm{mm}$, which is a tenth of the subaperture diameter. By calculating the Fourier transformation of the aperture function, equations (\ref{eq:nwf}) and (\ref{eq:dwf}) are numerically computed. The calculation of the weighting function is done for the layer height of every 100 m from 0 km to 50 km high. We set observation wavelength of $550\ \mathrm{nm}$ and zenith angle of $0\ \mathrm{deg}$. 
The computed weighting function matrices are shown in figure \ref{fig:WeightingFunction}. The two panels show the normal and differential weighting function matrices up to 20 km. The weighting function corresponding to each scintillation index is monotonically increasing as a function of height from the ground, meaning that scintillation becomes larger as the propagation distance increases. The order of the aperture patterns is sorted by the value of the weighting function at 20 km.

For turbulence profile reconstruction, we set a combination of fixed turbulence layer heights of 1, 2, 4, 8, and 16 km.
This is the same combination as MASS-DIMM except for the 0.5 km layer. 
The 0.5 km layer is removed because the 2 cm subaperture in our SH sensor is too large to constrain turbulence strength at less than $\sim 1$ km.
The smallest aperture of the MASS-DIMM instrument has a 1.3 cm diameter. 
The combination of heights is converted to propagation distance to the turbulence layer according to the elevation angle of the observation.
The columns of the weighting function matrices corresponding to the estimated propagation distances are extracted and we use them for the atmospheric turbulence profiling.

The turbulence profile is reconstructed by solving the inverse problem of equation (\ref{eq:nsi=wj}) and (\ref{eq:dsi=wj}) with respect to $C_n^2(h_i) \Delta h_i$. 
We use an iterative non-linear least square algorithm with some constraints to obtain reliable estimation. The cost function to be minimized is
\begin{align}
    \label{eq:minimization}
    \mathrm{minimize}\quad &\chi^2(\boldsymbol{J}) = \sum_m \frac{\{s_m - (\boldsymbol{W}\boldsymbol{J})_m\}^2}{\sigma_m^2} \\ \nonumber
    \mathrm{s.t.}\quad &\sum_i J_i < J_{\mathrm{slope}},
\end{align}
where $m$ is an index of NSI and DSI, $s$ and $\sigma$ mean the observed scintillation indices and their errors. 
$\boldsymbol{W}$ is the weighting function matrix and $\boldsymbol{J}$ is the turbulence profile, i.e. a vector of $C_n^2\Delta h$ for all the layer heights to be profiled. 
Therefore, modelled scintillation indices based on the turbulence profile, $\boldsymbol{W}\boldsymbol{J}$, are optimized to explain the observed scintillation indices within their measurement errors. 
As a constraint, we impose that the integrated strength of the turbulence profile should be smaller than the total turbulence strength, $J_{\mathrm{slope}}$, estimated by the slope auto-covariance analysis.
The condition corresponds to a physical requirement that turbulence strength traced by scintillation should be smaller than that traced with slope by ground layer strength, which cannot be traced by scintillation.
Also, we impose all the components of $\boldsymbol{J}$ to satisfy a condition of $-20<\mathrm{log}J_i\mathrm{(m^{1/3})}<-5$. 
This is an arbitrary condition in order to prevent the solution not to be negative, but the range is broad enough to include all possible atmospheric turbulence strength considering typical turbulence strength. 


\subsection{Scintillation temporal auto-covariance}
\label{sec:ana-scinticov}

We use the temporal auto-covariance maps of scintillation to obtain wind speed and direction profiles. 
Based on the frozen flow hypothesis, the scintillation pattern moves across the pupil with time following the wind speed and direction at altitudes where turbulence layers exist.
Therefore, wind speed and direction are estimated by examining each of the moving peaks on the temporal auto-covariance map of scintillation with a different time lag.
The strength of scintillation, which is defined as spot count time series normalized by its time average, depends on the turbulence strength and propagation distance from the turbulence layer, i.e. turbulence altitude.
Therefore, the auto-covariance map is more sensitive to turbulence at higher altitudes, but not sensitive to turbulence at ground, similar to the MASS technique.

To calibrate wind direction, a relationship between the direction of the subaperture arrangement on the primary mirror and the direction of the SH sensor field of view is first checked using a ray trace in the optical design. 
A relationship between the direction of the sensor field of view and the telescope field of view is checked from the telescope image rotator status.
Also, the wind direction measured in a plane parallel to the ground surface is obtained by converting the direction on the celestial plane to the horizontal coordinates.

\begin{figure*}
    \centering
    \includegraphics[width=\linewidth]{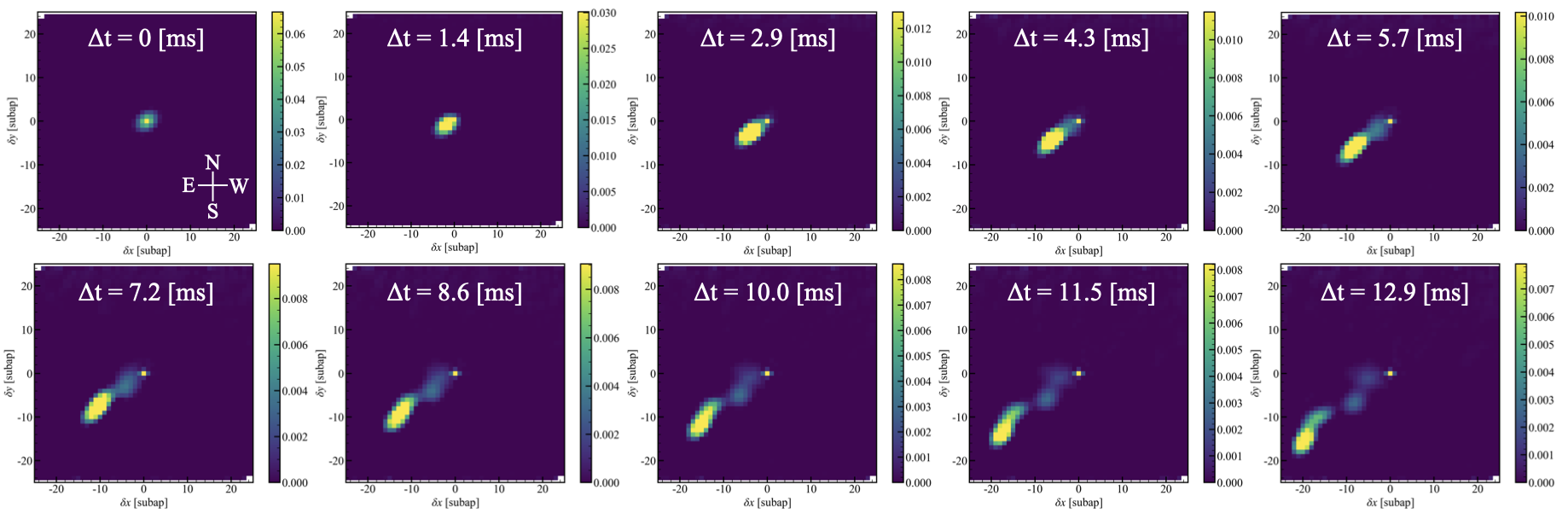}
    \caption[Example of scintillation auto-covariance map]{Example of scintillation auto-covariance maps for dataset taken at 02:42, Mar.14, 2023. From top-left to bottom-right, auto-covariance maps calculated with a time lag of 0-9 frames are arranged. The peak of the covariance signal moves toward the East-south direction as the time lag increases.}
    \label{fig:ScintillationAutocov}
\end{figure*}

\begin{figure}
    \centering
    \includegraphics[width=\columnwidth]{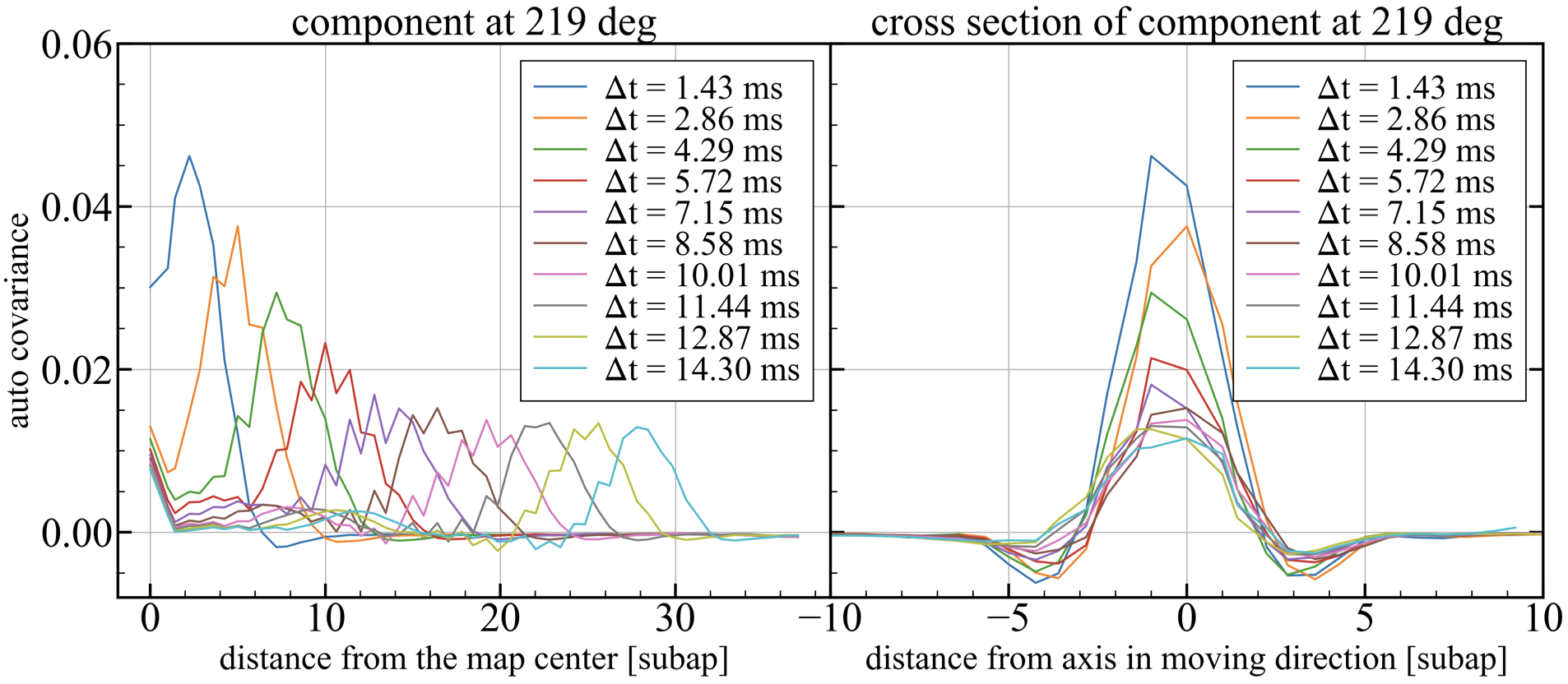}
    \caption[Example of scintillation auto-covariance analysis]{Example of scintillation auto-covariance analysis for the data taken at 02:42 on Mar.14, 2023. Left: cross-section of the covariance map along the direction of signal movement. The wind velocity of the signal component is estimated from the movement of the peak position at a constant speed. Right: Cross-section in the direction perpendicular to the direction of signal movement at the peak position of the signal. Wind height is estimated from the signal width since the width reflects the spatial scale of scintillation.}
    \label{fig:ScintillationAutocovAnalysis}
\end{figure}

The auto-covariance maps of scintillation obtained with increasing time lags are shown in figure \ref{fig:ScintillationAutocov}.
The several correlation peaks move toward the left-bottom direction on the map as the time lag increases, indicating that the turbulence in the upper layer is moving with the wind blowing from the northwest direction. 
The cross-section along the direction of signal movement is examined for each clump in each covariance map to find their peak position. 
The moving speed of the signal peak is converted to the wind speed. 
As an example, the cross-section of the brightest signal in figure \ref{fig:ScintillationAutocov} is shown in the left panel of figure \ref{fig:ScintillationAutocovAnalysis}. 
We correct an effect that the apparent wind speed is smaller by the factor of the cosine of observing the zenith angle, assuming that the real wind blows in a plane parallel to the ground.

Also, according to \citet{prieur2001scidar}, the FWHM of the scintillation auto-covariance signal in the Kolmogorov model is approximated as $\sqrt{\lambda z}$, where $\lambda$ is the observed wavelength and $z$ is the distance from the atmospheric turbulence layer to the ground. 
Therefore, it is also possible to estimate the height of the turbulence layer by analyzing the width of the auto-covariance signal.
We take a cross-section orthogonal to the moving direction at the location of each signal peak, which is shown in the right panel of figure \ref{fig:ScintillationAutocovAnalysis}. 
The width of the signal is measured by fitting them with the Gaussian function.
The procedure of measuring wind speed, direction and height is performed for several signal clumps found in each auto-covariance map. Clumps to be examined are selected based on that they have enough signal-to-noise ratio on the map and are independent of nearby clamps without overlaps. 

Furthermore, similar to the slope auto-covariance analysis, a method for obtaining an atmospheric turbulence profile by fitting the observed scintillation auto-covariance maps with the analytical ones is known as SCO-SLIDAR (\citealp{vedrenne2007c}). 
In this study, however, an atmospheric turbulence profile is estimated by SH-MASS, which is described in the next section.

\section{Results}
\subsection{Total Seeing}
\label{res:seeing}

\begin{figure}
    \centering
    \includegraphics[width=\columnwidth]{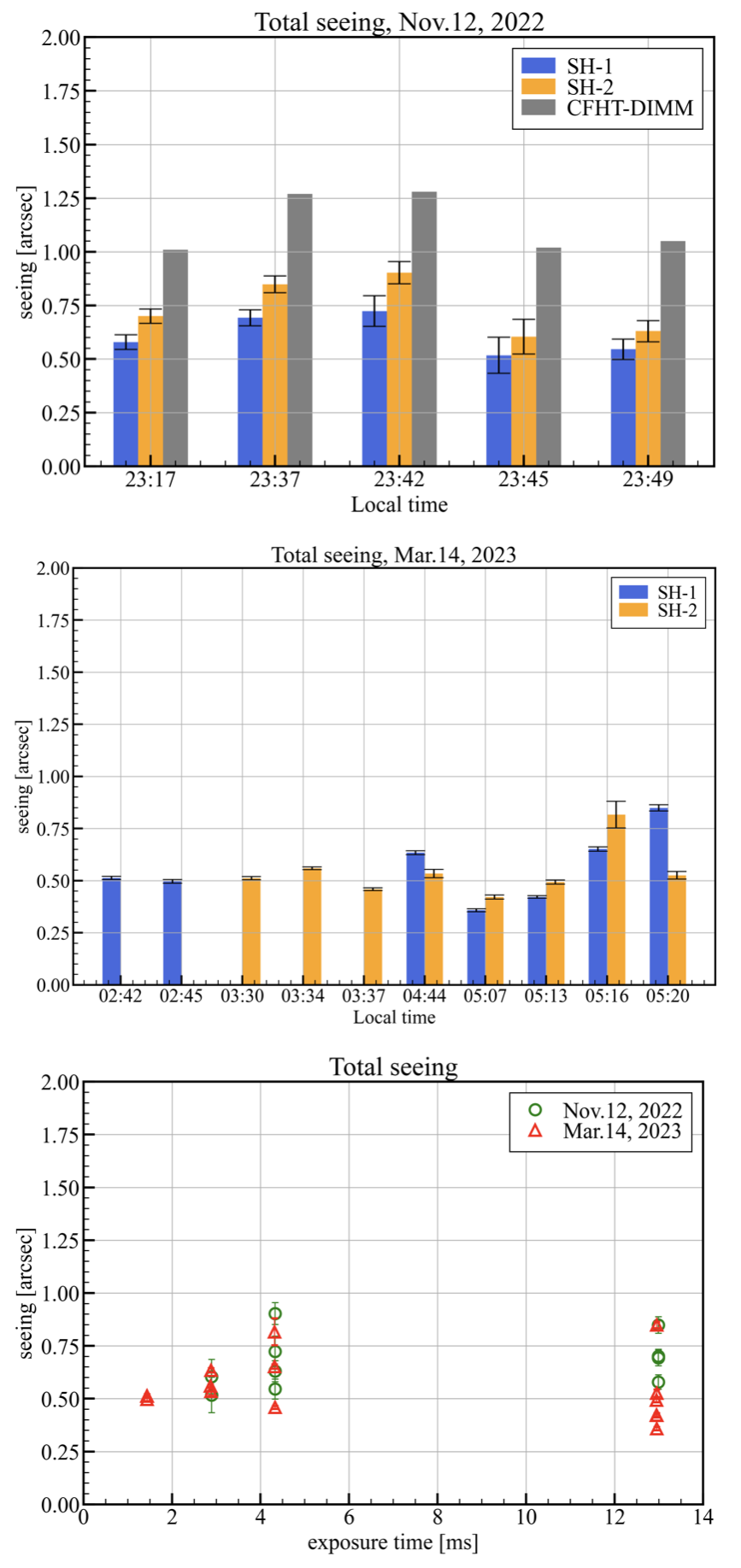}
    \caption[Total seeing measured by slope auto-covariance analysis]{Total seeing measured by slope auto-covariance analysis. Total seeing measured on Nov.12, 2022 and on Mar.14, 2023 are shown in the top and middle panels, respectively. Blue and orange bars represent results measured by SH sensors 1 and 2, respectively. The errorbar is coming from measurement error of auto-covariance map. The bottom panel shows the measured total seeing as a function of exposure time of SH sensors. The green and red symbols are measured on Nov.12, 2022 and Mar.14, 2023, respectively.}
    \label{fig:TotalSeeing}
\end{figure}

The results of the seeing estimation from the slope auto-covariance are shown in the top and middle panels in figure \ref{fig:TotalSeeing}. 
The histogram in the top and middle panels show the results of total seeing at each measurement time on Nov.12, 2022, and Mar.14, 2023, respectively.
The blue and orange bars indicate that the result is coming from the measurement of SH-1 and SH-2, respectively. 

The estimation error of the total seeing is propagated from the measurement error of the slope auto-covariance map, which is a 1-sigma error reflecting the signal-to-noise ratio of the SH sensor spots. 
The best signal-to-noise ratio is obtained at 02:42-03:37 on Mar.14 2023, when Arcturus with an apparent magnitude of $\sim 0$ is measured. 
At other times, the measurement error is not as small as the Arcturus measurement due to the faintness of the object and the exposure time of the imaging, but simultaneous measurements by the two SH sensors are achieved. 

The total seeing estimated independently from each SH sensor are in good agreement with each other within the estimation error, except for the end of the night, 05:16 and 05:20 on Mar.14, 2023, when some cloud covers the sky.
Based on the results, the medians of total seeing measured in our study are 0.661 arcsec on Nov.12, 2022 and 0.513 arcsec on Mar.14, 2023.

The bottom panel of figure \ref{fig:TotalSeeing} shows the measured seeing value as a function of the exposure time of the SH sensor. 
Seeing on both dates does not show a systematic trend depending on the exposure time, indicating that exposure time difference does not affect the measurements
Therefore, in our analysis, we do not correct any effect related to the sensor exposure time.

\subsection{Free atmospheric turbulence profile}
\label{res:turbulenceprofile}

\begin{figure*}
    \centering
    \includegraphics[width=\linewidth]{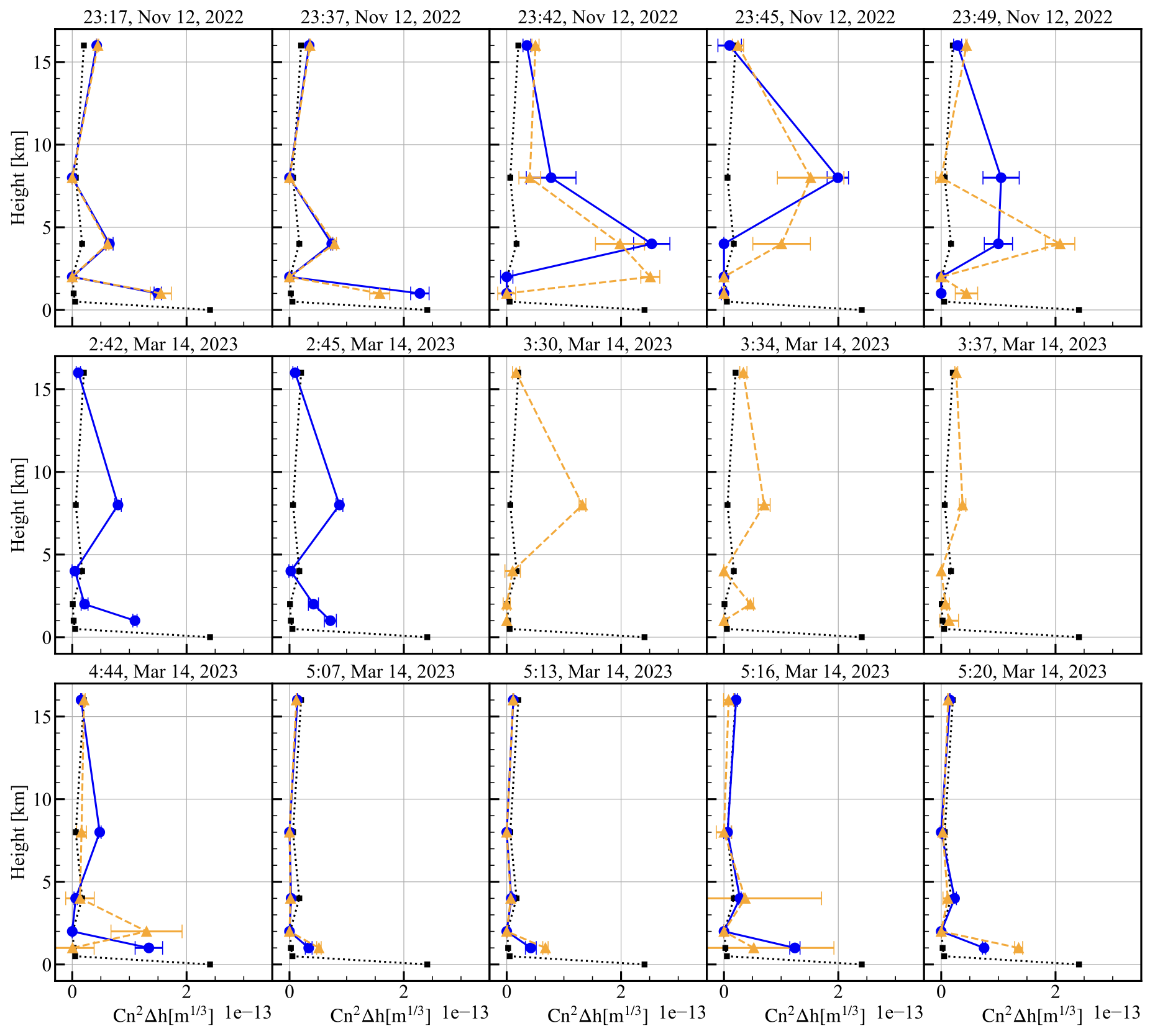}
    \caption[Free atmospheric turbulence profile]{Free atmospheric turbulence profile. 
    A reconstructed profile assuming 1, 2, 4, 8, and 16 km fixed height layers at each measurement time is shown in each panel. The blue and orange profiles are the results from SH-1 and SH-2, respectively. We plot the mean of simultaneous measurement by the four subdata on the two SH sensors. The errorbars represent standard deviation. The black dashed line is the median turbulence profile at Maunakea taken from \citet{els2009thirty} as a comparison.}
    \label{fig:FreeAtmosphereProfile}
\end{figure*}

The results of the free atmospheric turbulence profile reconstructed assuming the fixed layers at 1, 2, 4, 8, and 16 km are shown in figure \ref{fig:FreeAtmosphereProfile}.
Each panel shows an estimated profile at each measurement time.
The blue and orange profiles are the results of SH-1 and SH-2, respectively.
We plot the mean value of all four sub-datasets as symbols and standard deviation as errorbars. 
The black dotted profile stands for a representative profile obtained in the site testing campaign for the Thirty Meter Telescope (\citealp{els2009thirty}).
It shows median turbulence strength individually calculated for each height.

At most measurement times, the blue and orange profiles match and the error sizes are small enough to characterize turbulence strength at each altitude.
This indicates that the simultaneous measurements by the two SH sensors and the four sub-datasets show good agreement with one another.
As for the difference between profile shapes at 23:49, Nov.12, 2022, in which the blue profile shows its peak strength at 4 and 8 km while the orange profile has its peak at 4 km only, the orange profile should be more reliable according to the error size.
It can be thought that some of the sub-datasets in SH-1 show strong turbulence at 8 km instead of 4 km, and the mean of the four sub-datasets is affected.
The overall shape of the profile varies widely from time to time.
There is a strong layer at around 4 km on Nov.12, 2022 compared to the median profile of site testing.
However on Mar.14, 2023, turbulence at 1 km and 8 km are dominant components.

\begin{figure}
    \centering
    \includegraphics[width=\columnwidth]{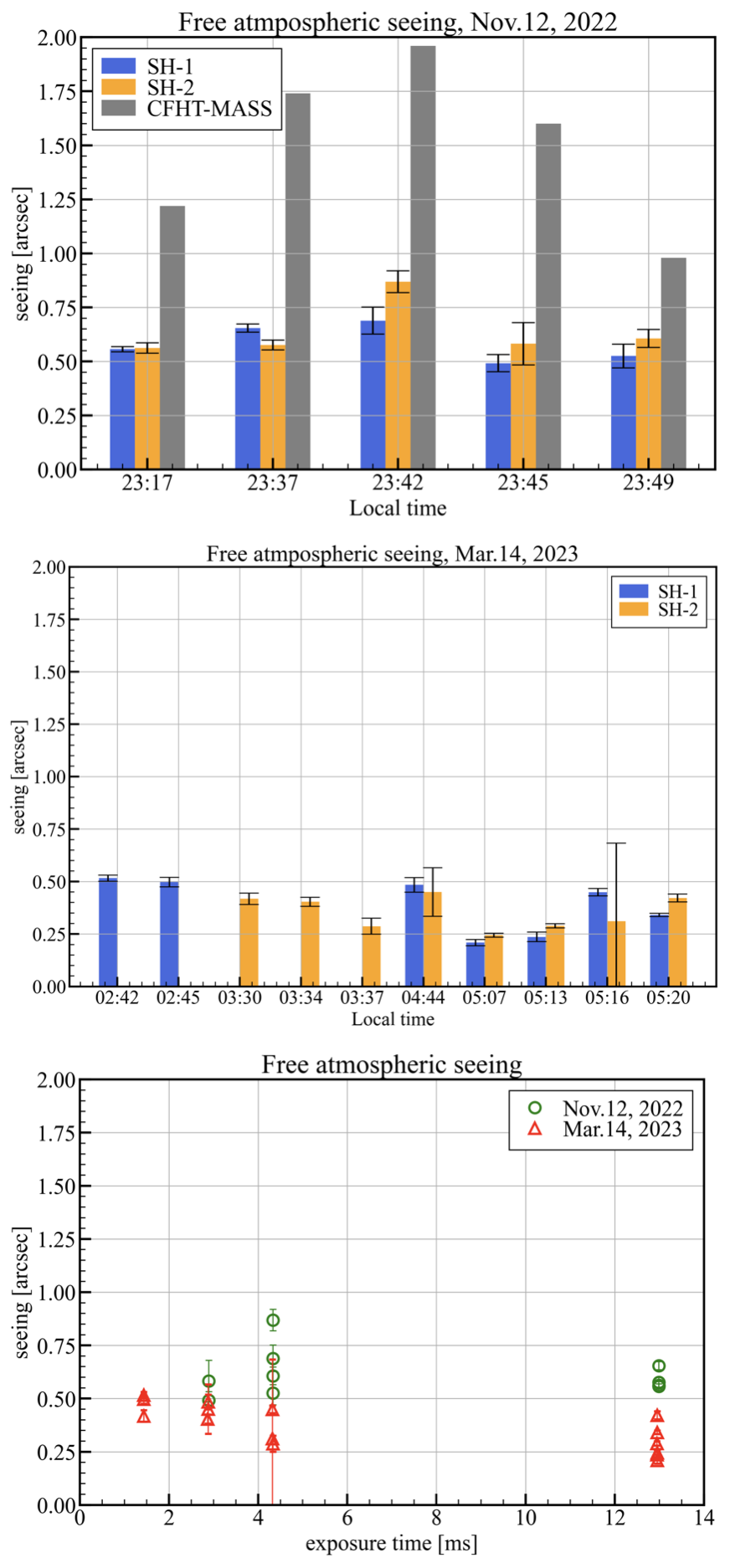}
    \caption[Free atmospheric seeing measured by SH-MASS]{Free atmospheric seeing measured by SH-MASS analysis. Free atmospheric seeing measured on Nov.12, 2022 and on Mar.14, 2023 are shown in the top and middle panels, respectively. Blue and orange bars represent results measured by SH sensors 1 and 2, respectively. The errorbar is coming from errorbars on the free atmospheric turbulence profile. The bottom panel shows the measured free atmospheric seeing as a function of the exposure time of SH sensors. The green and red symbols are measured on Nov.12, 2022 and Mar.14, 2023, respectively.}
    \label{fig:FreeAtmosphereSeeing}
\end{figure}

The top and middle panels of figure \ref{fig:FreeAtmosphereSeeing} show the integrated strength of the free atmospheric turbulence. 
The total strength is calculated as the sum of $C_n^2 \Delta h$ of the turbulence profile at each time shown in figure \ref{fig:FreeAtmosphereProfile}. 
The measurements by SH-1 and SH-2 are consistent within their error range between the two SH sensors.
The median of measured free atmospheric seeing is 0.579 arcsec for Nov.12, 2022 and 0.403 arcsec for Mar.14, 2023.

The bottom panel of figure \ref{fig:FreeAtmosphereSeeing} shows the free atmospheric seeing as a function of SH sensor exposure time.
Here also, measured seeing values do not show any trend depending on the exposure time.

\subsection{Wind profile}
\label{res:windprofile}

\begin{figure}
    \centering
    \includegraphics[width=\columnwidth]{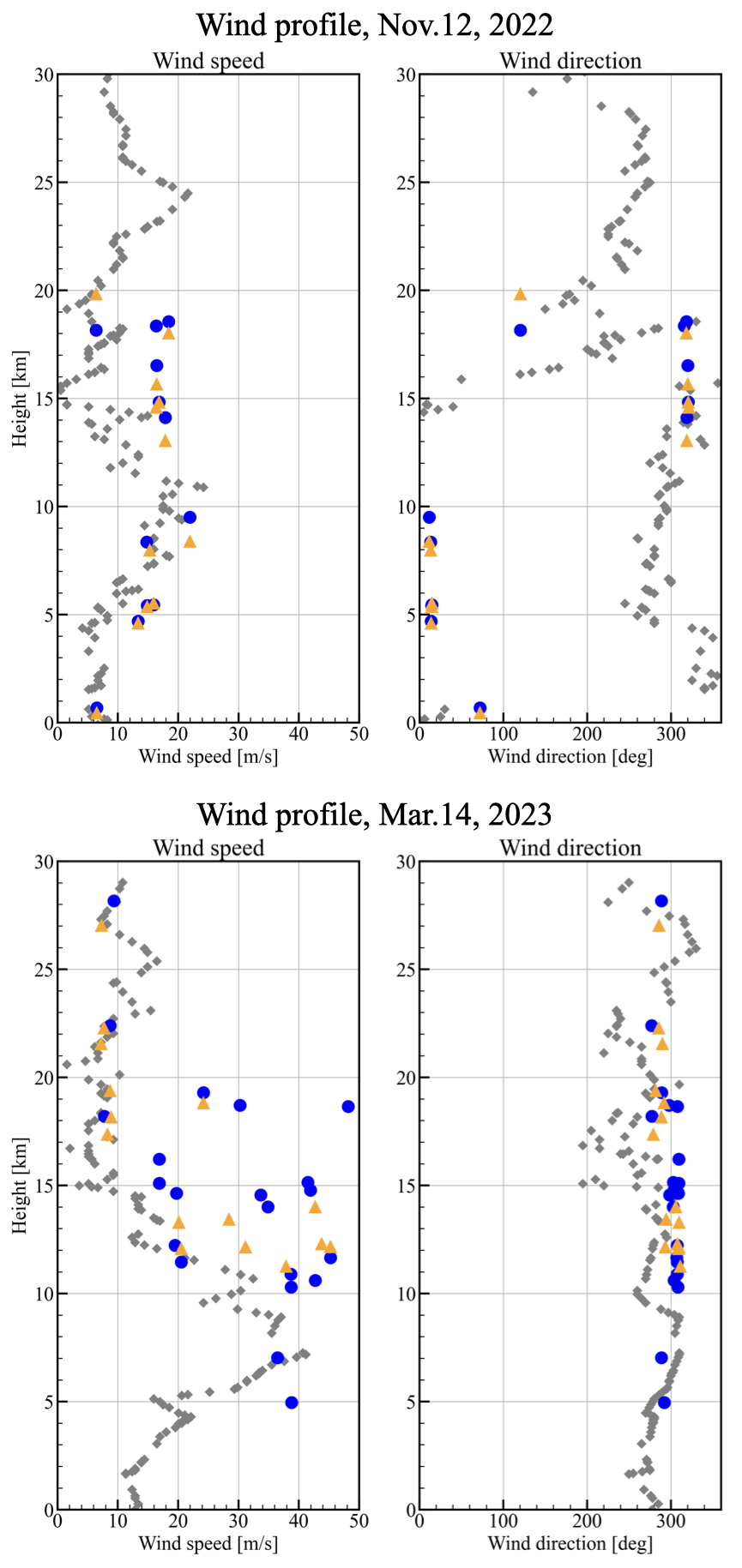}
    \caption[Wind speed profile]{Wind speed profile based on the scintillation auto-covariance analysis. Top panels show wind speed (left) and direction (right) on Nov.12, 2022 as a function of height from the telescope. Bottom panels are the same but for Mar.14, 2023. Each coloured symbol represents wind speed and direction measured with one of the moving peaks in the auto-covariance maps. Blue and orange symbols stand for measurement by SH-1 and SH-2, respectively. Results at all the measurement times are stacked in these figures. The grey dots are direct measurement results taken by a rawinsonde conducted around midnight on the same night at the Hilo airport.}
    \label{fig:WindSpeed}
\end{figure}

Wind profile, i.e. wind speed and direction as a function of height, is shown in figure \ref{fig:WindSpeed} as a result of the scintillation temporal auto-covariance analysis performed in section \ref{sec:ana-scinticov}.
The left and right panels are wind speed and direction profiles, respectively.
Results on Nov.12, 2022 are shown in the top panels and Mar.14, 2023 in the bottom.
The horizontal axis is the wind speed in the unit of $\mathrm{m \cdot s^{-1}}$ or wind direction in the unit of deg. 
The vertical axis means height measured from the altitude of the Subaru telescope (sea level of 4200 m).
The coloured points correspond to each clump seen in the auto-covariance map of each dataset. 
Blue and orange colours indicate that the measurement is coming from SH 1 and 2, respectively. 
The result is compared with the wind profile obtained by weather measurements, which is conducted by twice a day (noon and midnight) using rawinsonde data taken at the Hilo airport on Hawaii island, which is available from Maunakea Weather Center website (\url{http://mkwc.ifa.hawaii.edu/current/}).
The grey points show wind speed and direction by the rawinsonde at the nearest measurement time. 

By comparing the SHARPEST (coloured points) with the rawinsonde (grey points), the wind profiles measured by the two independent methods are in good agreement in spite of the measurement time difference. 
While rawinsonde is a direct measurement at all altitudes by a balloon, the scintillation auto-covariance is sensitive only to turbulence layers with strong scintillation intensity.
Therefore, the coloured points are not continuous in altitude direction and are distributed at different altitudes on the two measurement dates.
There are some points of discrepancy seen around the height of 5-10 km in the wind direction profile on Nov.12, 2022 or around the height of 20 km in the wind velocity profile on Mar.14, 2023. 
Whether this difference is simply due to differences in measurement time or due to systematic measurement errors that occur under certain conditions will need to be investigated.

\section{Discussion}

\subsection{Comparison with CFHT MASS-DIMM}
\label{sec:Dis_compwithCFHT}

Our turbulence measurement results are compared with simultaneous measurement of MASS-DIMM operated by the Canada-France-Hawaii Telescope (CFHT), which is available from the Maunakea Weather Center website (\url{http://mkwc.ifa.hawaii.edu/current/seeing/}).
The MASS-DIMM instrument is attached to a small telescope on a tower next to the CFHT and provides MASS turbulence profile at 0.5, 1, 2, 4, 8, 16 km and DIMM seeing every few minutes. 
In our observation run in November 2022, simultaneous measurements within 1 minute time lag with the CFHT MASS-DIMM are achieved. 
Unfortunately, the CFHT MASS-DIMM was not in operation during our observation run in March 2023 due to strong winds. 

In figure \ref{fig:TotalSeeing}, the grey histogram shows results from CFHT DIMM.
Its median value is 1.05 arcsec, a larger value than our measurements though they appear to be consistent in terms of trends in temporal change of strength.
This is partly because of the difference in ground layer turbulence at the two different sites.
More statistics on simultaneous measurements are needed in the future to investigate whether this difference is a systematic difference and what causes it.

Similarly, the grey histogram in figure \ref{fig:FreeAtmosphereSeeing} shows results from CFHT MASS.
The free atmospheric seeing by CFHT MASS, whose median value is 1.60 arcsec, is much larger compared to our measurements 
The median free atmospheric seeing by CFHT MASS is 1.60 arcsec, which is not only much larger than our measurements but also larger than the total seeing value by CFHT DIMM.
It is not a physical case given that the ground layer does not contribute to scintillation.
One of the reasons is the overestimated results of the CFHT MASS instrument.
This overestimation problem is known as MASS overshoot and common in the traditional MASS-DIMM instrument which estimates ground layer seeing from the difference between MASS seeing $\epsilon_{\mathrm{MASS}}$ and DIMM seeing $\epsilon_{\mathrm{DIMM}}$ as $(\epsilon_{\mathrm{DIMM}}^{5/3} - \epsilon_{\mathrm{MASS}}^{5/3})^{3/5}$.
According to \citet{tokovinin2007accurate}, the cause of the overshoot is that the actual atmospheric turbulence does not completely follow weak perturbation theory which is included in the assumption of MASS.
Some MASS-DIMM instruments are operated with the correcting algorithm empirically determined from numerical simulation, but such a correction has not been implemented for the CFHT MASS-DIMM yet.
In order to avoid the MASS overshoot, we put the seeing constraints when reproducing turbulence profiles as shown in equation (\ref{eq:minimization}).

\begin{figure}
    \centering
    \includegraphics[width=\columnwidth]{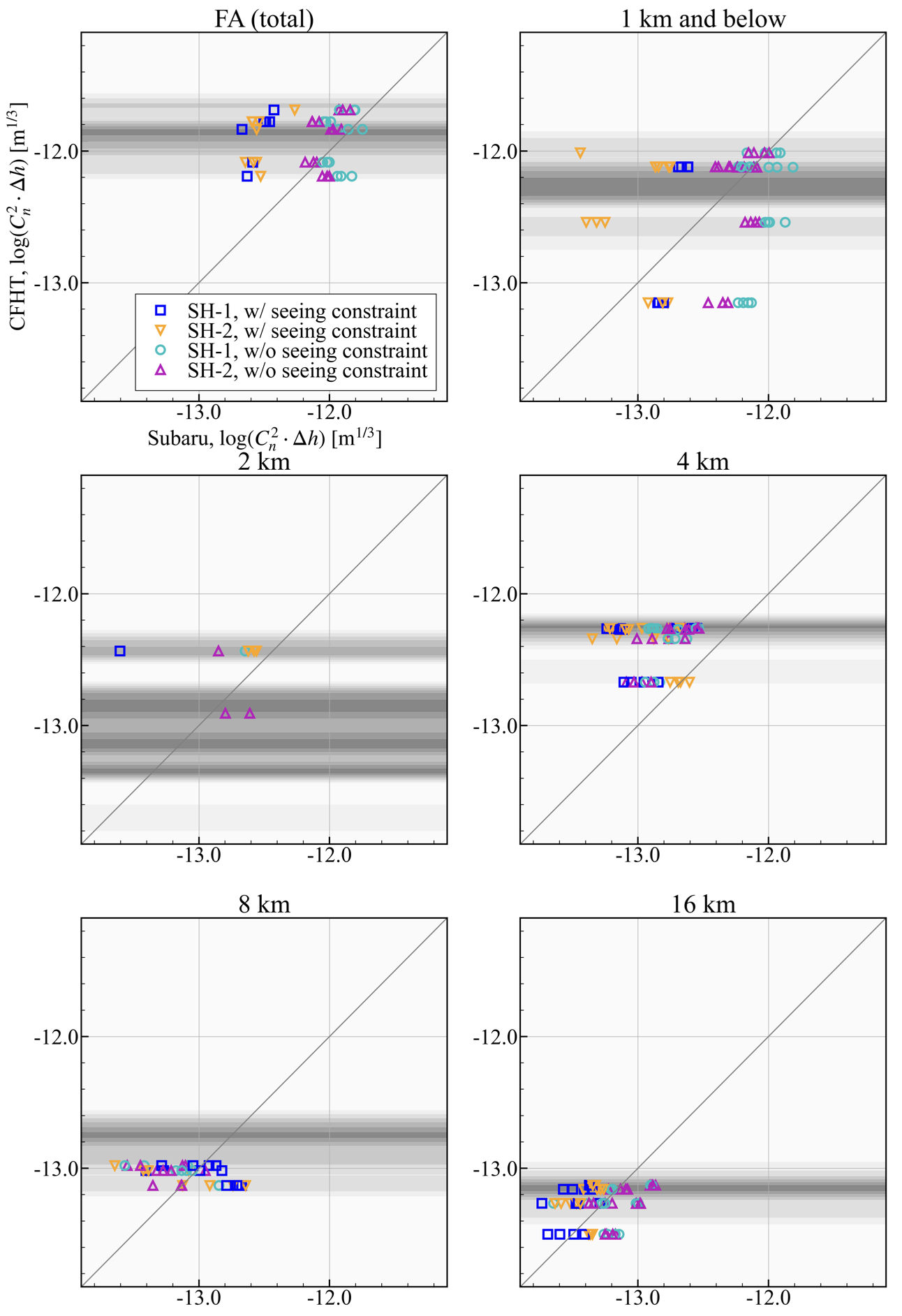}
    \caption[Comparison of free atmospheric turbulence measurements with CFHT MASS]{Comparison of free atmospheric turbulence measurement with CFHT MASS. Each panel shows a one-to-one plot of turbulence strength measured by our turbulence profiler and CFHT MASS. The temporal correspondence between the two measurements is less than 1 minute. As for the CFHT MASS measurement, which has higher temporal resolution, the probability distribution calculated from 34 measurements in 23:00-24:00 on Nov.12, 2022 is shown as gray shade.}
    \label{fig:ProfileCFHTcompare}
\end{figure}

Figure \ref{fig:ProfileCFHTcompare} shows a one-to-one plot of measurement results of CFHT MASS and SH-MASS. 
Each panel shows turbulence strength of a total, 1 km and below, 2 km, 4 km, 8 km, and 16 km as blue and orange symbols.
The sum of 0.5 km and 1 km strength is shown in the 1km-and-below panel for CFHT MASS measurement.
Since CFHT MASS has a higher temporal resolution of the measurement, data from the closest time is selected for comparison. 
The temporal difference between the two measurements is less than 1 minute.
The distribution of the CFHT MASS 34 measurements between 23:00 and 24:00 is shown in the grey area. 
From the comparison, it can be seen that SH-MASS measurements show good agreement with CFHT MASS at high layers such as 8 km and 16 km.
The 4 km turbulence of SH-MASS is slightly smaller than that of CFHT MASS.
As for the 2km turbulence, SH-MASS shows systematically much smaller turbulence strength than CFHT MASS.
However, according to the grey-shaded distribution, turbulence at the 2 km layer is smaller than in the surrounding layers, and measurements by CFHT MASS are not stable over time.
Therefore, both measurements are considered similar in behaviour.
SH-MASS results are smaller compared to CFHT MASS results in terms of the total and 1km-and-below turbulence, meaning that the large difference of free atmospheric seeing is coming from the difference of low-height estimation.

The cyan and magenta symbols, on the other hand, show the same one-to-one comparison but the SH-MASS profile is reconstructed without imposing constraints by total seeing measurement.
In other words, our SH-MASS estimation is performed under the same condition as CFHT MASS.
Based on the results, we can say that SH-MASS and CFHT MASS return consistent results at almost all altitudes.

\subsection{Consistency between scintillation auto-covariance map and SH-MASS}

In this study, wind speed and direction profiles are obtained by analyzing the scintillation auto-covariance map, and turbulence profiles are calculated based on the SH-MASS method. 
Because the information used in both analyses is identical in terms of spatial correlation of scintillation obtained with a single SH sensor, we estimate a turbulence profile by obtaining the signal intensity for each component on the scintillation auto-covariance maps shown in figure \ref{fig:ScintillationAutocov}.
The scintillation is quantified as a value normalized by the time-averaged value of brightness, i.e., the normal scintillation index. 
Therefore, the relationship between auto-covariance and turbulence strength is a weighting function assuming a single subaperture as the spatial filter. 

\begin{figure}
    \centering
    \includegraphics[width=\columnwidth]{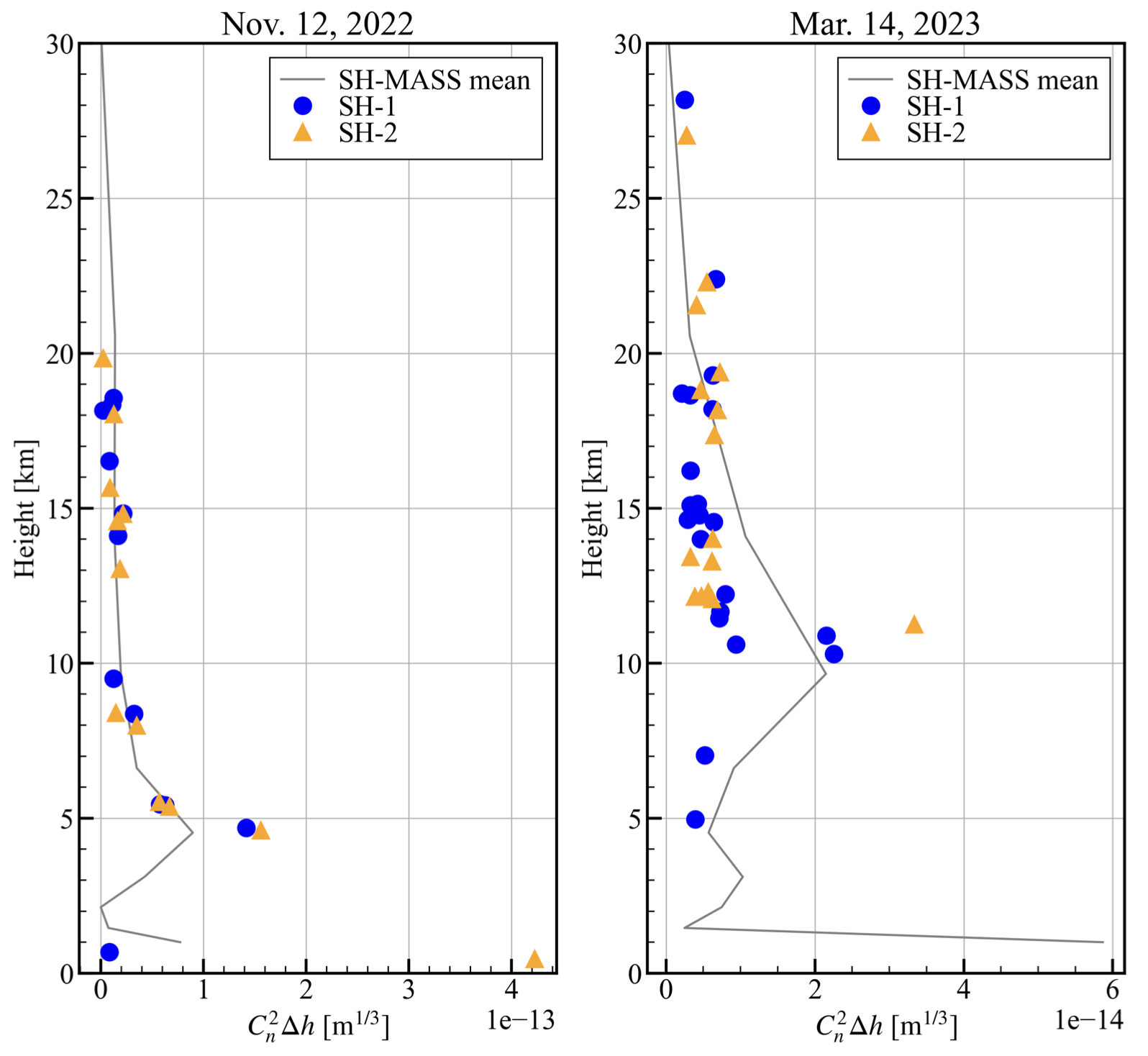}
    \caption{Turbulence strength profile based on the scintillation auto-covariance analysis. The left and right panels show $C_n^2 \Delta h$ as a function of height from the telescope measured on Nov.12, 2022 and Mar.14, 2023, respectively. Each coloured symbol represents turbulence strength measured with one of the moving peaks in the auto-covariance maps. Blue and orange symbols stand for measurement by SH-1 and SH-2, respectively. Results at all the measurement times are stacked in these figures. As a comparison, the mean of all the SH-MASS profiles is shown as grey lines. The SH-MASS profiles are reconstructed assuming 10 layers at 1.0, 1.5, 2.1, 3.1, 4.5, 6.6, 9.7, 14.1, 20.6, and 30.0 km}
    \label{fig:WindStrength}
\end{figure}

The turbulence profiles obtained by the analysis are shown in figure \ref{fig:WindStrength}.
As a comparison, we show the mean of all SH-MASS profiles on each day as grey lines. 
The SH-MASS profiles are reconstructed assuming that there are 10 turbulence layers between 1 and 30 km.
In the Nov.12, 2022 profiles, high turbulence intensity of around 1 km and local intensity peaks around 4-5 km are seen. 
In the Mar. 14, 2023 profile, there is a localized area of strong turbulence around 10-11 km.
These characteristics are in good agreement with the trend seen in the SH-MASS profile. 

\section{Conclusion}

In this article, we report the details of the SHARPEST, an atmospheric turbulence profiling project at the Subaru telescope and its initial results.
The purpose of the project is to obtain (1) free atmospheric turbulence profiles up to $\sim 20$ km with an altitude resolution of a few km and (2) ground layer turbulence profiles below $\sim 1$ km with a fine resolution of a few tens of meters, as information for LTAO and GLAO projects at the Subaru telescope, respectively.

In order to achieve the goal, we develop a turbulence profiler consisting of two SH sensors with a fine subaperture of 2 cm and perform engineering observations of star pairs brighter than a magnitude of $\sim 6$ in the $V$ or $R$ band with a separation angle of 3-5 arcmin.

For measurements of a single SH sensor, we apply SH-MASS, a method for free atmospheric turbulence at higher than 1 km with the same or higher resolution of a few km than classical MASS based on spatial correlation of scintillation.
In addition to the profiling of turbulence strength, we measure total seeing from the spatial correlation of wavefront and profiles of moving speed and direction of turbulence from the temporal correlation of scintillation.
We also perform SLODAR, a method for ground layer turbulence below $\sim 400$ m with a high resolution of $\sim 20$ m from the correlation of wavefront measurements in two different directions, and the results will be presented in the next paper.

As a result, the obtained total seeing shows consistent values between simultaneous measurements by the two SH sensors, but systematically smaller values compared to CFHT DIMM by $\sim 37 \%$ in median.
The reason for the difference might come from the difference in ground layer turbulence including dome seeing between the Subaru and CFHT sites. 

The obtained free atmospheric turbulence profiles at 1, 2, 4, 8, and 16 km altitudes are compared with the previous TMT site test campaign and measurement by CFHT MASS. The profiles measured simultaneously by the two independent SH sensors and four sub-datasets show consistent results. The profiles show different shapes between the two observation runs, supporting that the distribution of turbulence is distinguished by SH-MASS. Compared to CFHT MASS, our SH-MASS results show a much smaller total turbulence strength possibly because we impose a physical constraint that the total turbulence strength measured by SH-MASS should be smaller than the total seeing while CFHT MASS is not constrained by the CFHT DIMM measurements and suffers from overshoot problem (total turbulence strength measured by CFHT MASS is larger than that measured by CFHT DIMM). If we perform SH-MASS without physical constraint, the results of SH-MASS and CFHT MASS show good agreement.

We also reconstructed atmospheric turbulence profiles based on the strength and width of the scintillation auto-covariance signals. The comparison with the mean free atmospheric turbulence profile by the SH-MASS shows consistent results regarding the heights where the profiles have a peak intensity.

The wind speed and direction profile are compared with results from the rawinsonde, which is measured using a balloon.
Although our observation and the rawinsonde measurements are not simultaneous due to the low temporal frequency of the rawinsonde, our measurement of wind speed and direction shows very good agreement with the directly measured results by the rawinsonde.

Through the study, we establish the method to constrain free atmospheric turbulence profiles, measure the total seeing values, and reconstruct the wind profiles by analysing data from a single SH sensor with a high spatial sampling.


\section*{Acknowledgements}
The authors thank the staff of the Subaru telescope for much support for the development, installation, and observations of the turbulence profiler.
This work is financially supported by a grant from the Joint Development Research supported by the Research Coordination Committee, National Astronomical Observatory of Japan (NAOJ), National Institutes of Natural Sciences (NINS),
and by the FY2021 supplementary budget from the Japanese Ministry of Education, Culture, Sports, Science and Technology (MEXT).
The work is also supported by the Japan Society for the Promotion of Science (JSPS) KAKENHI Grant Numbers JP17H06129, JP21H05583 and JP22J14095, and by a grant from the Hayakawa Satio Fund awarded by the Astronomical Society of Japan.
HO is supported by the JSPS as a young research fellow and the Graduate Program on Physics for the Universe (GP-PU), Tohoku University as a research assistant.

\section*{Data availability}
The data underlying this article will be shared on reasonable request to the corresponding author.



\bibliographystyle{mnras}
\bibliography{bibliography} 








\bsp	
\label{lastpage}
\end{document}